\documentclass[prb,showpacs,twocolumn,superscriptaddress]{revtex4-1}
\usepackage[utf8]{inputenc}
\usepackage[american,british]{babel}
\usepackage[T1]{fontenc}
\usepackage[pdftex]{graphicx}
\usepackage{graphicx, xcolor}
\usepackage{dcolumn}
\usepackage{bm}
\usepackage{amsmath,amsthm,amssymb}
\usepackage{amsmath}
\usepackage{color}
\usepackage{verbatim}
\usepackage{ulem}
\usepackage{amssymb,mathtools}
\usepackage{float}
\usepackage{tikz,calc}

\usetikzlibrary{shapes, arrows,positioning,fit,automata}
\tikzstyle{plain} = [draw,thick,circle,inner sep=0,minimum size=0.5cm,font=\footnotesize]
\tikzstyle{mps} = [draw,thick,rectangle,rounded corners=.1cm,inner sep=0,minimum size=0.5cm]
\tikzstyle{mpo} = [draw,thick,circle,inner sep=0,minimum size=0.5cm]
\tikzstyle{index} = [-,thick,font=\footnotesize]
\tikzstyle{virtual} = [-,thick,dotted,font=\footnotesize]
\tikzstyle{site} = [draw,solid,circle,minimum size=2pt,inner sep=0pt,outer sep=0pt,fill=black]

\def \tu {0.18cm}

\definecolor{darkGreen}{RGB}{0,110,0}
\definecolor{darkBlue}{RGB}{0,0,130}
\usepackage[colorlinks,citecolor=darkGreen,linkcolor=darkBlue,%
urlcolor=blue,hyperindex]{hyperref}

\newcommand{\bra}[1]{\left\langle #1 \right|}
\newcommand{\ket}[1]{\left| #1 \right\rangle}
\newcommand{\ave}[1]{\left< #1 \right>}
\newcommand{\braket}[2]{\langle #1 | #2 \rangle}
\newcommand{\brakett}[3]{\langle #1 |  #2 | #3 \rangle}

\renewcommand{\(}{\left(}
\renewcommand{\)}{\right)}
\renewcommand{\[}{\left[}
\renewcommand{\]}{\right]}
\renewcommand{\H}{\mathcal{H}}

\newcommand{\be}{\begin{equation}}
\newcommand{\ee}{\end{equation}}

\begin{document}
\title{Diagnosing Potts criticality and two-stage melting in one-dimensional hard-boson models}

\author{G. Giudici}
\thanks{These authors contributed equally to this work.}
\affiliation{The Abdus Salam International Centre for Theoretical Physics, strada Costiera 11, 34151 Trieste, Italy}
\affiliation{SISSA, via Bonomea 265, 34136 Trieste, Italy}
\affiliation{INFN, via Bonomea 265, 34136 Trieste, Italy}
\author{A. Angelone}
\thanks{These authors contributed equally to this work.}
\affiliation{The Abdus Salam International Centre for Theoretical Physics, strada Costiera 11, 34151 Trieste, Italy}
\affiliation{SISSA, via Bonomea 265, 34136 Trieste, Italy}
\author{G. Magnifico}
\affiliation{The Abdus Salam International Centre for Theoretical Physics, strada Costiera 11, 34151 Trieste, Italy}
\affiliation{Dipartimento di Fisica e Astronomia dell'Universit\`a di Bologna, I-40127 Bologna, Italy}
\affiliation{INFN, Sezione di Bologna, I-40127 Bologna, Italy}
\author{Z. Zeng}
\affiliation{School of Physics, Sun Yat-sen University, Guangzhou 510275, China.}
\author{G. Giudice}
\affiliation{Max-Planck Institute of Quantum Optics, Hans-Kopfermann-Str. 1, 85748 Garching, Germany}
\author{T. Mendes-Santos}
\affiliation{The Abdus Salam International Centre for Theoretical Physics, strada Costiera 11, 34151 Trieste, Italy}
\author{M. Dalmonte}
\affiliation{The Abdus Salam International Centre for Theoretical Physics, strada Costiera 11, 34151 Trieste, Italy}
\affiliation{SISSA, via Bonomea 265, 34136 Trieste, Italy}

\date{\today}

\begin{abstract}
We investigate a model of hard-core bosons with infinitely repulsive nearest-
  and next-nearest-neighbor interactions in one dimension, introduced by
  Fendley, Sengupta and Sachdev in Phys. Rev. B 69, 075106 (2004). Using a
  combination of exact diagonalization, tensor network, and quantum Monte Carlo
  simulations, we show how an intermediate incommensurate phase separates a
  crystalline and a disordered phase. We base our analysis on a variety of
  diagnostics, including entanglement measures, fidelity susceptibility,
  correlation functions, and spectral properties. According to theoretical
  expectations, the disordered-to-incommensurate-phase transition point is
  compatible with Berezinskii-Kosterlitz-Thouless universal behavior. The
  second transition is instead non-relativistic, with dynamical critical
  exponent $z> 1$. For the sake of comparison, we illustrate how some of the
  techniques applied here work at the Potts critical point present in the phase
  diagram of the model for finite next-nearest-neighbor repulsion. This latter
  application also allows us to quantitatively estimate which system sizes are
  needed to match the conformal field theory spectra with experiments
  performing level spectroscopy. 
\end{abstract}

\maketitle

\section{Introduction}

Recent years have witnessed considerable experimental progress aimed at
realizing and manipulating atomic physics systems with long-range
interactions~\cite{Lahaye2009, Baranov2012a}.  Examples of this span a variety
of platforms, including trapped ions~\cite{blatt2012quantum}, cold polar
molecules~\cite{Carr2009rev}, and atomic gases of strongly dipolar atoms such
as Cr, Dy, and Er~\cite{Lahaye2009}.  Strong nonlocal interactions can also be
induced on neutral atoms by coupling their atomic ground states to Rydberg
states\cite{Saffmann2010, Low2012, Lim2013, Jones2017, Morsch:aa}.  The large
dipole moments displayed by the latter allow the engineering of large dipolar and
van der Waals interactions~\cite{Leseleuc:aa, Lukin2017, Zeiher2016, Jau2015,
Faoro:2016aa, Keesling:aa}, which offer the possibility of performing quantum
simulation of long-range, strongly interacting systems~\cite{Bloch2012}.

These new avenues of experimental realization have caused renewed interest in
many theoretical models which, besides displaying remarkable physical
phenomena, might be realized within present experimental settings.  In this
work, we will focus on one such model, first introduced in
Ref.~\onlinecite{Fendley2004} by Fendley, Sengupta, and Sachdev (FSS). The model
describes an array of one-dimensional strongly interacting hard-core bosons in
the presence of occupation constraints on nearest-neighbor (NN), and with
additional interactions on next-to-nearest-neighbor (NNN), sites. Initially
discussed due to its connections with integrable models, its successful
implementation in Rydberg atom arrays~\cite{Lesanovsky2012, Lukin2017} has
driven further theoretical investigation~\cite{Ghosh:2018aa, Samajdar2018,
Chepiga2018, Whitsitt:2018aa, Chepiga:aa}.

\begin{figure}
  {\hspace{-2mm}\includegraphics[scale=0.33]{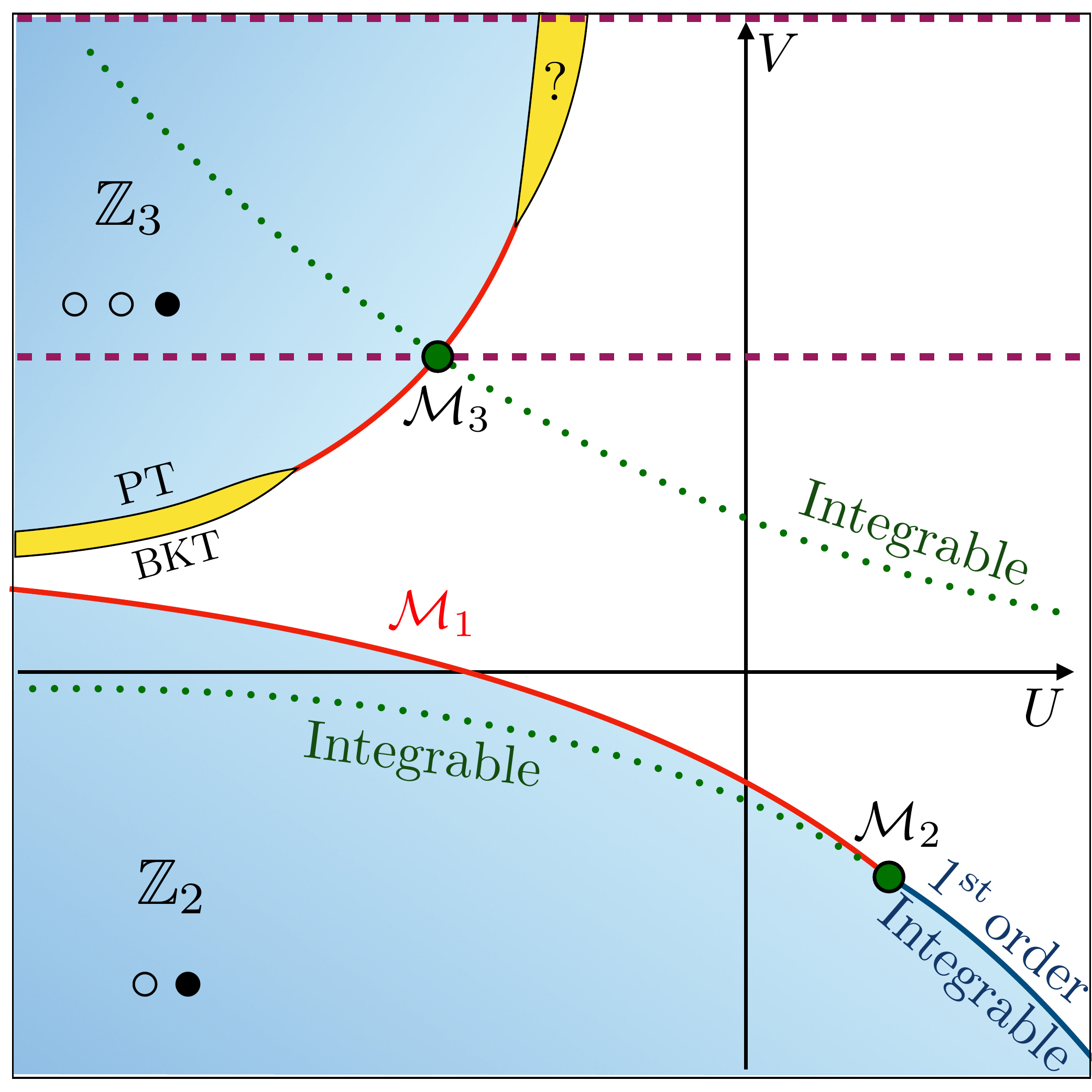}}
  \caption{Phase diagram of the model Hamiltonian Eq.~\eqref{ham1}. Ordered
  phases are colored in light blue. Red (blue) lines indicate second (first)
  order phase transitions. On the green dotted lines the model is integrable.
  The integrable line in the lower half plane is on top of the transition line
  when the transition is of the first order. The two lines separate at the
  tricritical point $\mathcal{M}_2$, where the transition becomes continuous.
  After this point, the second-order phase transition belongs to the Ising
  universality class. The integrable line in the upper half plane crosses the
  second-order transition line exactly at the $\mathcal{M}_3$ critical point,
  belonging to the Potts model universality class. Below this point, on the
  transition line, a gapless phase (lower yellow region, not in scale) opens,
  enclosed within a Japaridze-Nersesyan-Pokrovsky-Talapov (JNPT) and Berezinskii-Kosterlitz-Thouless
  (BKT) transition.  Above this point, the opening of a gapless phase (upper
  yellow region, not in scale) is under debate. Purple dashed lines are studied
  in this work.}
  \label{phase_diag}
\end{figure}

The phase diagram of the FSS model displays a variety of phases and phase
transitions which still needs to be fully understood (see
Fig.~\ref{phase_diag}). In particular, there are two ordered phases with
$\mathbb{Z}_2$ and $\mathbb{Z}_3$ order~\cite{Fendley2004} and a disordered
phase in which long-range correlations can be incommensurate with the lattice
spacing~\cite{Chepiga2018}. The transition separating the period-two ordered
phase from the disordered one is well understood: one switches from first to
second order at a tricritical point, the second-order line belonging to the
Ising model universality class~\cite{DiFrancesco1997}. The phase diagram hosts
an integrable line, which crosses the boundary between the
$\mathbb{Z}_3$--ordered and disordered phases at a critical point belonging to
the Potts universality class~\cite{Henkel_1999, DiFrancesco1997}. Below it, the
phase transition is still continuous, but Lorentz invariance is broken at low
energy by an irrelevant chiral perturbation which changes the critical
exponents \cite{Huse1984, Chepiga2018}. Eventually, a gapless phase opens on the
line and the order-disorder transition becomes a Luttinger liquid
phase~\cite{gogolin_book, DiFrancesco1997} with incommensurate long-range
correlations\cite{Fendley2004, Chepiga2018}. 

Above the Potts critical point the situation is more controversial, and so far
several scenarios have been proposed. The chiral perturbation {\it i)} might
become relevant, making the transition first order; {\it ii)} it might lead to
the same effect as below the Potts point, thus leaving the transition
continuous; or {\it iii)} it might stabilize an intervening gapless phase
between the ordered and disordered phase. In principle, the transition may
remain in the Potts universality class; however, this scenario is unlikely as
it would require fine-tuning. Recently, numerical evidence has been provided
both in favor of a continuous phase transition surviving on the whole
line~\cite{Samajdar2018} and in favor of a gapless phase opening at a point
above the Potts one~\cite{Chepiga2018}, while simulations on small system sizes
are compatible with Potts universality extending up to infinite NNN
repulsion~\cite{Ghosh:2018aa}.

The aim of this paper is to clarify the nature of the
$\mathbb{Z}_3$-order-to-disorder transition above the integrable line. We focus
on the regime of infinite NNN repulsion, which we refer to as the
\textit{doubly blockaded regime}, in analogy with the more common NN blockade.
The reason for this choice is threefold: it being the farthest regime from the
Potts critical point, it may allow for a comparatively larger incommensurate
phase (if any) thanks to the fact that the role of perturbations moving away
from the exactly solvable line (see Fig.~\ref{phase_diag}) is typically larger;
it is of easy experimental access; it is amenable to exact simulations up to
comparatively larger sizes with respect to the rest of the phase diagram.

We show that the melting of the ordered phase, at the boundary of the phase
diagram, takes place via an intermediate gapless phase. This critical phase is
enclosed between two continuous phase transitions. From the disordered side, the
transition is of the BKT type, while from the ordered side the universality
class is not captured by conformal field theory (CFT). 

We compute many of the critical exponents of these transitions with different
methods. As we discuss below, our findings are only able to provide a lower
bound for the size of the incommensurate (IC) phase, due to the presence of
anomalously large finite-size effects; small sizes systematically reduce the
size of the IC phase.  In parallel, we test some of the methods employed on the
exactly located Potts critical point; this helps us to emphasize differences
and similarities between the two melting phase transitions. We also give a full
characterization of the Potts critical point by computing its critical
exponents, and by matching the low-lying energy spectrum on the lattice with
the universal predictions provided by conformal field theory. This
characterization provides a quantitative and unambiguous testbed to verify
Potts quantum criticality in experiments based on spectroscopic probes.

We employ various methods to tackle the problem numerically, focusing on
periodic geometries in order to avoid boundary effects, which are particularly
detrimental for constrained models in the vicinity of ordered phases. We
exploit at best the small quantum dimension of the Hilbert space to compute the
ground state and the lowest excited states exactly up to $54$ sites. We perform
studies of up to 120 sites via quantum Monte Carlo (QMC), using an
imaginary-time path integral method sharing many similarities with the worm
algorithm~\cite{Prokofev1998-1}, adapted to simulate Hamiltonians with
off-diagonal terms such as those of the FSS model and with updates designed to
automatically respect its occupation constraints. We use the density matrix
renormalization group (DMRG) algorithm~\cite{White1992} to compute the ground
state of periodic chains up to $108$ sites. In this case, we implement the
constraint by giving a large penalty to the states which are not allowed in the
Hilbert space. We also present results for the experimentally realized open
chain scenario by simulating open chains up to $718$ sites with a 1-site DMRG
algorithm formulated in the matrix product state (MPS) language, which allows us
to realize the constraint exactly by representing efficiently the global
projector on the constrained Hilbert space as a matrix product operator (MPO). 

The structure of the paper is as follows. In Sec.~\ref{sec2}, we present the
Hamiltonian of the model, stressing its importance in relation to Rydberg atom
experiments~\cite{Lukin2017} and reviewing in detail previous theoretical results. In
Sec.~\ref{sec3}, we discuss the methods we employ, and investigate the vicinity of the Potts transition point, in particular, performing an analysis based on level spectroscopy. In
Sec.~\ref{sec4}, we study in detail the doubly blockaded regime. In Sec.~\ref{sec5}, we draw our conclusions and discuss some future perspectives.

\section{Model Hamiltonian and review of previous results}
\label{sec2}

\subsection{Hard-core interactions in Rydberg-blockaded gases}

Alkali atoms in Rydberg states~\cite{Gallagher, Low2012, Lim2013, Jones2017} are
characterized by one of their electrons occupying an energy level with high
principal quantum number ($n > 40$). Several features of the atomic state are
strongly influenced by this type of excitation. Examples include a very
long radiative lifetime $\tau \sim n^3$, a large orbital radius $r_o \sim n^2$,
and a strong dipole moment $d \sim n^2$: in the case of rubidium excited to the
$50p$ state, these quantities can assume values on the order of $10^2$ $\mu$s,
$10^{-1}$ $\mu$m, and $10^3$ $ea_0$, respectively.

These remarkable characteristics determine the strong interactions
between Rydberg atoms. The dominant contribution to the latter will
be of the dipole-dipole [$V(r) \sim r^{-3}$ with the interatomic distance $r$]
or van der Waals ($V(r) \sim r^{-6}$) type in the presence and absence of
external polarizing field, respectively, due to the aforementioned strong
dipole moments. The large values displayed by the latter cause these
interactions to be very strong, with a typical scaling for the van der Waals
coupling constant being $C \sim n^{11}$.

Due to their extremely high values, Rydberg--Rydberg interactions far outstrip
any other energy scale in the system, and therefore play a fundamental role in
determining the behavior of a many-body Rydberg ensemble. One of the most
remarkable phenomena induced in this interaction-dominated picture is the
Rydberg blockade~\cite{Saffmann2010, Lukin2001}, in which the excited atomic
levels, due to the strength of the interatomic potential $V(r)$, are heavily
shifted from their noninteracting positions. Since $V(r)$ becomes stronger at
short distances, a Rydberg atom creates an effective ``exclusion zone'' around
itself, where no other excitations to the Rydberg state can take place (due to
the excited level being too strongly off-resonance with respect to the
excitation mechanism, usually a laser in experimental setups).

In a many-body system of Rydberg atoms, this phenomenon can be used to simulate
an effective hard-shell constraint: gauging the experimental parameters
(namely, the frequency and the detuning of the laser used to induce excitations
to the Rydberg state), the blockade radius can be tuned to simulate occupancy
constraints for sites at different distances on an optical lattice, reproducing
the most significant feature of models such as the FSS Hamiltonian, as realized in Ref.~\onlinecite{Lukin2017}.

\subsection{Hard-boson model phase diagram}

The Hamiltonian of the FSS model is given by
\be \label{ham1}
H = -  \sum_i ( d_{ \mathstrut i } + d^\dagger_{ \mathstrut i } ) + U  \sum_i
n_{\mathstrut i } + V  \sum_i n_{\mathstrut i }  n_{\mathstrut i +2  }  
\ee
where $d^\dagger_i$ ($d_i$) is the creation (destruction) operator for a
hard-core boson on site $i$ and $n_i = d^{\dagger}_i d_i$. The Hilbert space is
subjected to the constraint $n_i n_{i+1} = 0$; namely, two particles cannot
occupy NN sites.  When this restriction is imposed, the number of states
$\mathrm{dim} \H_L $ in the Hilbert space for a chain of length $L$ satisfies,
in the case of open boundary conditions (OBCs), the recursive equation
\be
\mathrm{dim} \H_L = \mathrm{dim} \H_{L-1} + \mathrm{dim} \H_{L-2}
\ee
whose solution is the Fibonacci sequence, which behaves asymptotically for
large $L$ as $\mathrm{dim} \H_{L} \sim \phi^L$, where $\phi = 1.6180...$ is the
golden ratio. The dimension of $\H$ becomes even smaller in the limit $V \to
\infty$, which is equivalent to saying that there have to be at least two empty
lattice sites between two particles, i.e., $n_i n_{i+1} = 0$ and $n_i n_{i+2} =
0$. It is easy to see that in this case $\mathrm{dim} \H_L $ satisfies the
equation
\be
\mathrm{dim} \H_L = \mathrm{dim} \H_{L-1} + \mathrm{dim} \H_{L-3}
\ee
which asymptotically means $\mathrm{dim} \H_L \sim \zeta^L$, with $\zeta =
1.4655..$. 

The model was first proposed as the quantum version of the 2-dimensional
classical hard-square model~\cite{Fendley2004}, which is known to host two
integrable lines~\cite{Baxter2007}. One of the two lines crosses the
period-three-to-disorder line exactly at the Potts critical point, whose
location is thus known analytically. The classical-to-quantum mapping results
in a constrained quantum Hilbert space which is not in product form. As already
noted in Ref.~\onlinecite{Chepiga2018} and further discussed below, the
peculiar way order is realized in the system causes extremely strong
finite-size effects, especially when OBCs are applied.
This poses challenges for tensor-network based techniques, which usually rely
on these boundary conditions, since the computational effort must be increased
in order to access larger system sizes.  Oppositely, the milder scaling of the
Hilbert space dimension allows us to exactly diagonalize the system up to lengths
which roughly double the usual lengths accessible in spin chains.  Since
periodic boundary conditions (PBCs) eliminate boundary effects, in addition to
providing momentum symmetry for a direct diagonalization of the quantum
Hamiltonian, they will be employed throughout this work, with the exception of
the tensor network simulations presented in Sec.~\ref{sec4}. 

The phase diagram of the model is depicted in Fig.~\ref{phase_diag}. The two
integrable lines are parametrized by
\be \label{param}
V ( U + V ) = 1.
\ee
One of the two lines is defined on the upper half plane $(U,V)$, and crosses
the order-disorder transition line exactly at the Potts critical point
mentioned above, for $V = V_c = [( \sqrt{5} + 1 )/2 ]^{5/2}$; it is thus
described at low energies by the third conformal field theory in the minimal
series, $\mathcal{M}_3$~\cite{DiFrancesco1997}. The gapped ordered phase
extends to a region in the quadrant $V>0$, $U<0$, where the order-disorder
transition is not always a sharp transition. In particular, it was shown in
Ref.~\onlinecite{Fendley2004} that in the limit case $U \to -\infty$ and $V = -
U/3$ the separation line is in fact a thin gapless phase (yellow region in
Fig.~\ref{phase_diag}) characterized by the Luttinger liquid (LL) universality
class. The transition from the ordered phase to the gapless phase belongs to
the Japardze-Nersesyan-Pokrovsky-Talapov (JNPT) universality class~\cite{japaridze1978,PT1979}, and has dynamical
critical exponent $z=2$. Conformal invariance is then restored in the continuum
description and the transition from the gapless phase to the disordered phase
is of the Berezinskii-Kosterlitz-Thouless type~\cite{gogolin_book}. Moreover, a
recent detailed analysis\cite{Chepiga2018} exhibited strong numerical evidence
that the very same picture persists on the order-disorder transition line, up
to a Lifshitz point located below the integrable line, beyond which the
transition is sharp and of the chiral Huse-Fisher type~\cite{Huse1984}. However,
the precise location of the Lifshitz point could not be estimated. The authors
also confirmed the position and nature of the Potts critical point by computing
the correlation length critical exponent $\nu$ coming from both phases.

What happens above the integrable line is more controversial. A DMRG-OBCs
study~\cite{Chepiga2018} is in favor of a chiral transition up to another
Lifshitz point after which a LL phase opens again, with a PT transition on the
ordered side and a BKT transition on the disordered side. The width of the
intervening LL phase was estimated at the order of 0.001. It was also noted
that, above the Potts point, boundary effects are sizable at system sizes on
the order of several hundred sites, as testified by an anomalous scaling of the
von Neumann entropy. Instead, an exact diagonalization (ED) study~\cite{Samajdar2018}, using PBCs,
indicated that there is no Lifshitz point, and the transition remains chiral up
to $V=\infty$, with a dynamical critical exponent $1 < z \lesssim 1.33$.

In what follows we will focus on two lines at constant $V$ (purple dashed lines
in Fig.~\ref{phase_diag}). The phase diagram on the first line is very well
understood and we will use it as a benchmark to test field theory predictions
in this exotic quantum chain. The second line is located at $V=\infty$ and, as
discussed above, its phase structure is still under debate.

\section{Potts critical point}
\label{sec3}

In this section, we study the finite-size properties of the Potts critical
point. This is important not only to test some of the methods we are going to
employ in the following sections, but also to understand which universal
properties can be experimentally measured with the available setups of
$\simeq$50 spins. Moreover, it is of theoretical interest, as there are very
few lattice realizations of Potts criticality that can be studied in such a
systematic fashion~\cite{pfeuty1981, Rittenberg1987, Rittenberg1986}.

The CFT behind the Potts model universality class is one of the modular
invariant realizations of the third model in the minimal series:
$\mathcal{M}_3$~\cite{Dotsenko1984}. Its central charge is $4/5$ and the most
relevant primary fields, namely the energy density and the order parameter,
carry anomalous dimensions $\eta_\varepsilon = 4/5$ and $\eta_\sigma = 4/15$.
These two numbers imply that the correlation length and order parameter
critical exponents are $\nu = (2 - \eta_\varepsilon)^{-1} = 5/6$ and $\beta =
\nu \eta_\sigma/2 = 1/9$.

The position of the Potts critical point in the phase diagram of the quantum
Hamiltonian in Eq.~\eqref{ham1} is known exactly by integrability
arguments~\cite{Fendley2004} and its location has been checked numerically both
via gap scaling analysis~\cite{Samajdar2018} as well as from vanishing inverse
correlation length\cite{Chepiga2018}. Its critical exponents have been computed
on the lattice, and a clear signature of the underlying CFT has been
observed~\cite{Fendley2004,Samajdar2018,Chepiga2018}. However, the low-energy
spectrum of the lattice Hamiltonian has never been matched with the CFT one and
a full characterization of the phase transition has never been given.
Furthermore, contrary to the lattice Potts model, the $\mathbb{Z}_3$ symmetry
is not an exact global symmetry of the FSS model. It is thus non-trivial to
identify the whole operator content from the energy eigenvalues on the lattice.

Before performing level spectroscopy, we test some of the methods we will
employ in the next section to witness second-order phase transitions without
any assumption on the spacetime symmetry of the underlying field theory, namely
nonanalyticity in the quantum concurrence~\cite{Fazio2002,Fazio2008} --- which
is a measure of single spin entanglement --- as well as in the fidelity
susceptibility~\cite{Gu:2008aa,Zanardi:2006aa}. The latter also allows us to
extract the critical exponent $\nu$ of the ordered
phase~\cite{You2007,Capponi2009,Capponi2010}. We then compute the central charge
of the CFT from the logarithmic scaling of the entanglement
entropy~\cite{Calabrese2009} and we show that the CFT regime is reached with
system sizes accessible to present experiments. We proceed by matching momentum
symmetry sectors on the lattice with $\mathbb{Z}_3$ sectors in the CFT. We
match several low-lying eigenvalues with the corresponding primary fields and
we discuss the finite-size scaling corrections with respect to CFT predictions. Finally,
we extract the anomalous dimension $\eta_\sigma$ of the order parameter by
comparing its lattice two-point function with the one of a CFT on a
ring~\cite{DiFrancesco1997}.

\subsection{Critical point location}
\label{concfid_potts}

As we will see below, in order to locate the critical point, it is useful to
utilize a procedure which is not biased by any assumption on the nature of the
phase transition, such as conformal invariance and a consequent scaling of the
gap with a dynamical critical exponent $z=1$.  Here we use two methods based on
the nonanalytic behavior displayed by generic functions in the presence of
continuous phase transitions. The concurrence is a measure of entanglement for
spin systems~\cite{Fazio2008,Plenio:2007aa}, and is defined as
\be \label{conc}
C  = \mathrm{max} ( 0 , \lambda_1 - \lambda_2  - \lambda_3  - \lambda_4 )
\ee
where the $\lambda_i$ are the square roots of the eigenvalues in decreasing
order of the matrix $ \sqrt{\rho_{i,j}} (\sigma^y \otimes \sigma^y )
\rho_{i,j}^{*} (\sigma^y \otimes \sigma^y ) \sqrt{\rho_{i,j}}$, where
$\rho_{i,j}$ is the reduced density matrix of two sites located at positions
$i$ and $j$ (here we show results for $j=i+2$) \footnote{Since the Hilbert
space is not a product of single-site Hilbert spaces, in order to compute the
reduced density matrix we simply plunge back the ground state of the system
into the full Hilbert $ (\mathbb{C}^2)^{\otimes L}$.}. The function $C(U)$ is
expected to have an infinite derivative at a gapless critical point in the
thermodynamic limit~\cite{Fazio2002,Fazio2008}.  At finite size, the derivative
$\partial_U C$ has a peak which sharpens with increasing system size at a value
$U^*(L)$ which converges to the critical point when $L \to \infty$.

\begin{figure}
{\includegraphics[scale=0.5]{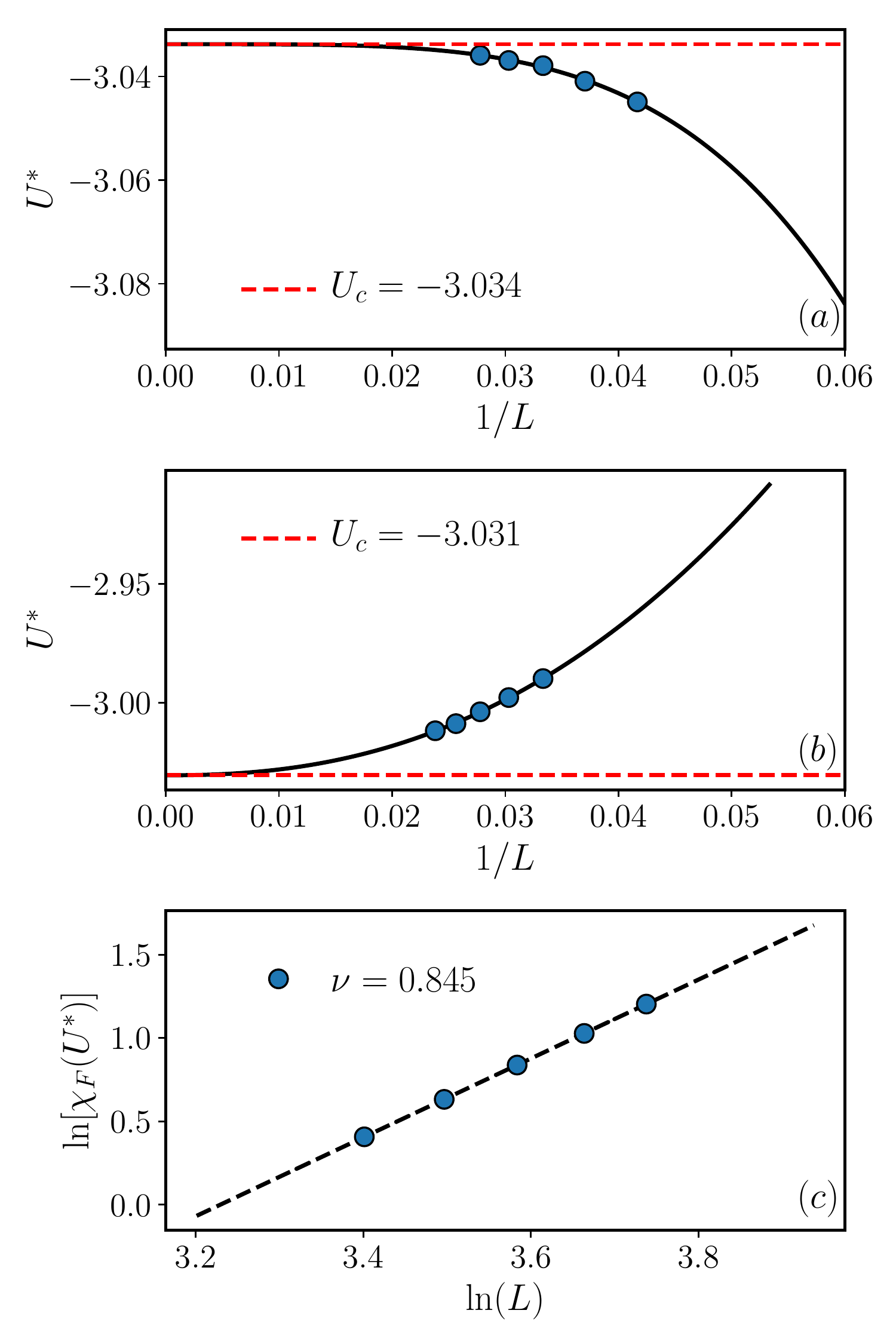}\hspace{2mm}}
\caption{(a) Power-law fit of the peak position $U^*(L)$ of the first
  derivative of the concurrence Eq.~\eqref{conc}, for $L$ from 24 to 36 sites.
  The scaling exponent extracted from this range of system sizes is $\gamma =
  4.0 \pm 0.1$, but it is not stable including smaller sizes. The critical
  position we get from the fit is $U_c = -3.03 \pm 0.04$. (b) Power-law fit of
  the peak position $U^*(L)$ in the fidelity susceptibility
  Eq.~\eqref{fid_susc}, for $L$ from 30 to 42 sites. The result is stable when
  smaller system sizes are included. Taking into account small variations
  with respect to the range of lengths employed in the fit, the scaling exponent and
  the critical point position we get are $\gamma = 2.4 \pm 0.1$ and $U_c =
  -3.03 \pm 0.01$. (c) Scaling of the maximum of $\chi_F$ according to
  Eq.~\eqref{fid_scal} for $L$ from 30 to 42 sites. The correlation length
  critical exponent slightly increases when smaller system sizes are included
  in the fit. By taking into account variations with respect to the range of lengths
  fitted we get $\nu = 0.84 \pm 0.01$, in good agreement with the exact value
  $\nu = 5/6 = 0.8333..$.}
\label{fid_conc} 
\end{figure}

In Fig.~\ref{fid_conc}(a), we plot the value of the position of the peak of
$\partial_U C$, $U^*(L)$, at the Potts critical point as a function of $1/L$.
The position of the critical point at $L \to \infty$, $U_c$, is obtained by
fitting $U^*(L)$ with the power-law function, $U^*(L) = U_c + A/L^{\gamma}$.
The best-fitting exponent for system sizes from 24 to 36 sites is $\gamma = 4.0
\pm 0.1$ and the extracted position of the critical point is $U_c = -3.03 \pm
0.04$, in good agreement with the exact value $U_c = -3.0299\dots$.  Note
however, that the result is not stable when smaller system sizes are included in
the fit. We attribute this instability to the limited number of sizes we can
reliably simulate in the scaling regime, due to the challenging nature of the
calculation of concurrency.

Another quantity that is used to locate and characterize the critical point is
the fidelity susceptibility
\be \label{fid_susc}
\chi_F = \frac{ - 2 \ln | \braket{\psi_0(U) }{ \psi_0(U + \delta U ) } | }{ \delta U^2 }
\ee
where $|\psi_0(U)\rangle$ is the ground-state wave function for a fixed value
of $U$. As the derivative of the concurrence, $\chi_F$ exhibits a peak at the
position $U^{*}(L)$, when plotted as a function of $U$. The size scaling of
$U^{*}(L)$ provides an alternative approach to establish the position of the
critical point, $U_c$; see Fig.~\ref{fid_conc}(b). In contrast to the
concurrence, the numerical calculation of the fidelity is less expensive and
allows us to reach system sizes up to $L=42$. This yields a best-fitting result
which is stable against the range of system sizes included in the fit for $L\ge
24$. The best-fitting parameters we get considering lengths from 27 up to 42
sites gives $U_c = -3.03(1)$, where the error takes into account variations
against the system sizes included in the fit. Furthermore, a scaling theory for
the height of the peak of $\chi_F$ does exist\cite{Capponi2009, Capponi2010}
and allows us to obtain the correlation length critical exponent via
\be \label{fid_scal}
\chi_F( U^* ) \sim L^{2/\nu} .
\ee
Note that this power-law scaling is independent of the value of the dynamical
critical exponent $z$. In this way we get a value of $\nu$ in perfect
agreement with the expected value for the Potts model universality class; see
Fig.~\ref{fid_conc}(c).

\begin{figure}
{\includegraphics[scale=0.5]{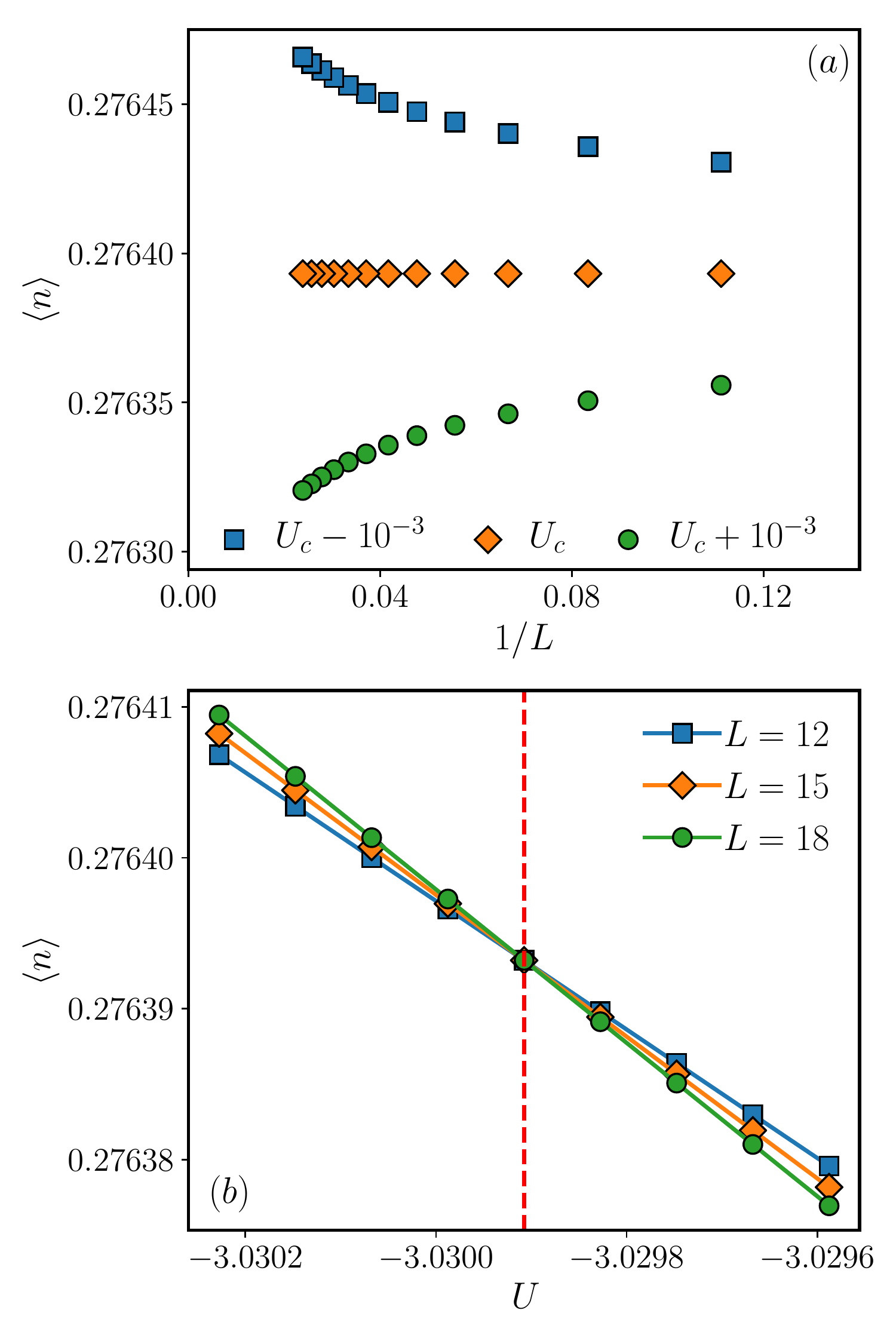}\hspace{2mm}}
\caption{(a) Finite-size scaling of the boson density at the transition point
  and close to it. Exactly at the transition point, the density does not scale.
  (b) Boson density as a function of $U$ for $V=V_c$ and different chain
  lengths. The lines sharply cross at the transition point (dashed red line)
  for any system size, since finite-size corrections vanish exactly at the
  critical point.}
\label{density}
\end{figure}

Finally, we wish to mention a very peculiar fact which allows us to locate the
critical point with arbitrary precision and arbitrary small system sizes:
exactly at the critical point, the on-site boson density has vanishing
finite-size corrections. The position of $U_c$ can thus be obtained by
measuring the boson density for different system sizes and tuning the couplings
until size independence is observed. We believe that this fact is due to the
integrable structure beyond the critical spin chain. In Fig.~\ref{density}, we
report the finite-size scaling of the density at the critical point and for two
values of $U$ very close to it, together with the curve crossing of densities
computed for different system sizes as a function of $U$ for $V = V_c$, which
allows a precise determination of the position of the critical $U$.

\begin{figure}
{\includegraphics[scale=0.5]{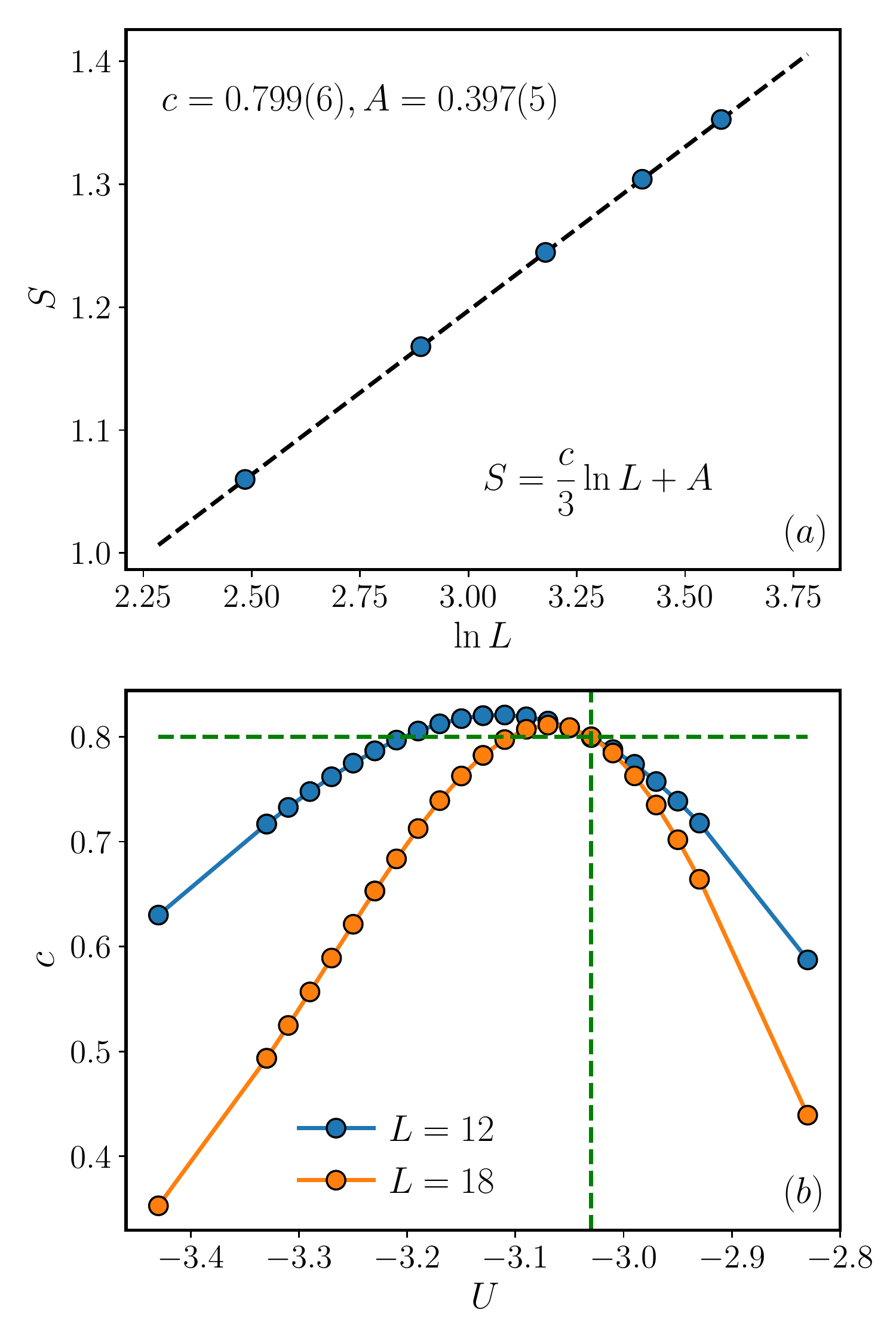}\hspace{2mm}}
\caption{(a) Finite-size scaling of the entanglement entropy of a
  half partition, for $L$ from $12$ to $36$ sites. The slope is the one
  expected from CFT already for system sizes $L \le 24$, indicating negligible
  finite-size corrections to the CFT predictions Eq.~\eqref{ee_scal}. (b)
  Effective central charge as defined in Eq.~\eqref{effc}. The peak is
  sharpening as the system size is increased and the peak position is moving
  towards the expected value $U_c = -3.0299...$. Curves for different lengths
  cross almost exactly at this value of $U$, indicating the presence of a
  single critical point in which the effective central charge is
  nondecreasing.}
\label{ee}
\end{figure}

\subsection{Entanglement entropy}

Continuous, relativistic phase transitions in a 1D system display a logarithmic
divergence of the entanglement entropy. Once conformal invariance is ensured,
an inexpensive way to identify the universality class is by computing the
coefficient of the logarithmic growth of the entanglement as a function of the
subsystem size. This coefficient is known to be proportional to the central
charge of the CFT~\cite{Calabrese2009}, and for the case of half partition in
PBCs reads
\be  \label{ee_scal}
S = \frac{c}{3} \ln L + A.
\ee
In Fig.~\ref{ee}(a), we plot the entanglement entropy for the critical values
$(U_c,V_c)$ analytically known. This result shows how moderate sizes are
already yielding a very precise value for the central charge. In
Fig.~\ref{ee}(b), the effective central charge, defined
as~\cite{Laeuchli:2008aa}
\be \label{effc}
c = 3 \, \frac{ S( 2 L ) - S(L) }{ \ln 2 },
\ee
is plotted for fixed $V = V_c$, and varying $U$ across the transition. The
central charge exhibits a bell-shape dependence on $U$, observed in other cases
as well~\cite{Dalmonte:2015aa}, with a peak which is approaching the expected
position marked with a green dashed line. Note that different bells touch only
at the critical point, which is the only value of $U$ at which the effective
central charge is not decreasing with increasing system size. This is in
agreement with Zamolodchikov's theorem~\cite{zamo1986} in the presence of a
single critical point.

\subsection{CFT level spectroscopy}

\begin{figure}
{\includegraphics[scale=0.5]{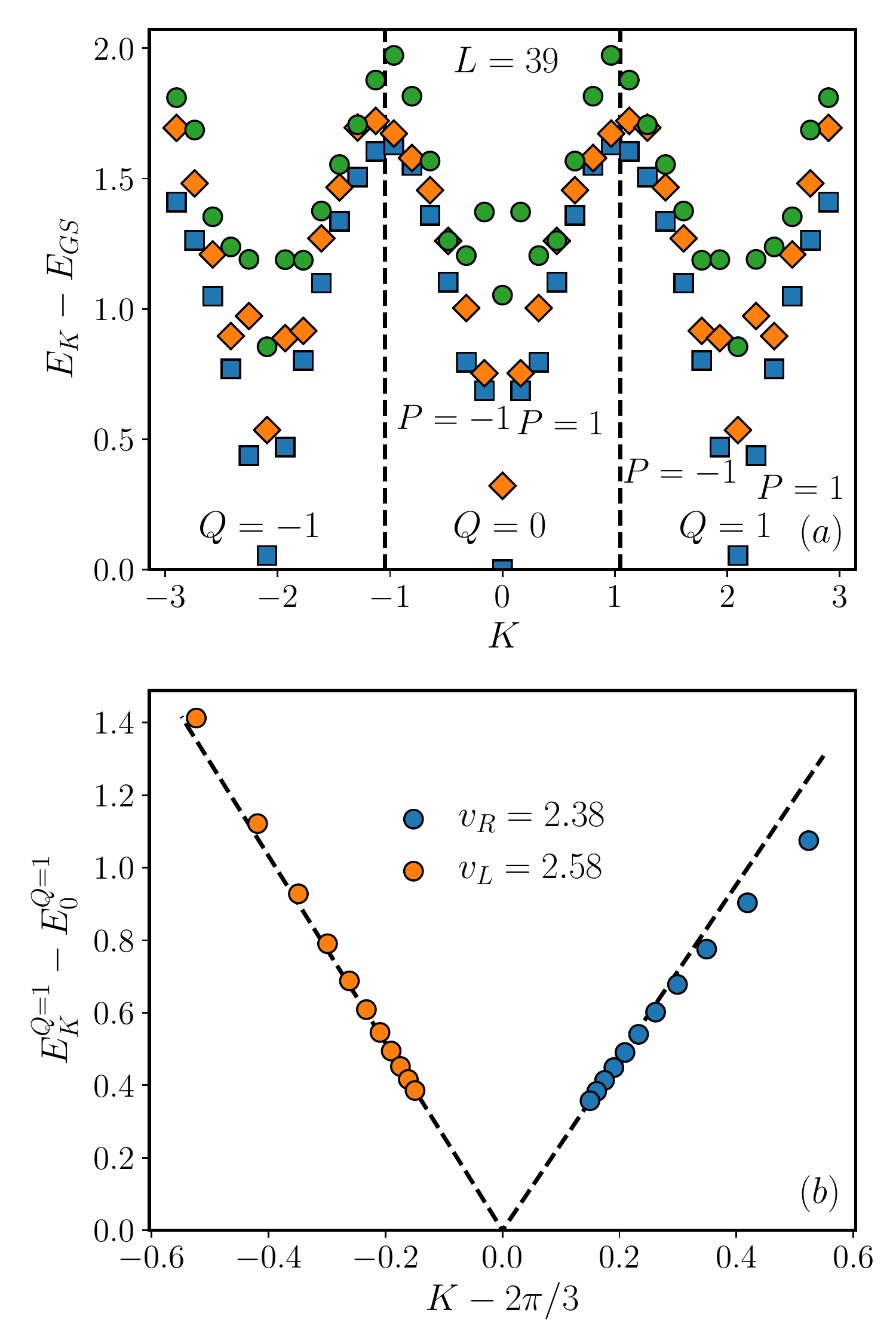}\hspace{2mm}}
\caption{(a) Lowest-lying eigenvalues of lattice momentum sectors for a chain
  of length $L=39$. Eigenvalues close to $K=0$ and $K=\pm 2 \pi /3$ correspond
  to the $\mathbb{Z}_3$ sectors $Q=0$ and $Q=\pm 1$ in the CFT.  Conformal
  towers are already distinguishable and primary operators corresponding to
  each energy level can be easily guessed by comparing the lowest gaps with
  Eq.~\eqref{cft_gaps} together with Eqs.\,\eqref{op0} and \eqref{op1}. (b)
  Linear fitting of the lowest eigenvalue close to the ground state of the
  $Q=1$ sector for different system sizes from $L=12$ to $L=42$.  The zero
  reference energy is taken as the ground-state energy of the sector for the
  given system size. Chiral symmetry is broken on the lattice, most likely
  because of an irrelevant perturbation which scales away in the thermodynamic
  limit.}
\label{velocity}
\end{figure}

Computing the entanglement entropy is a convenient way of extracting universal
information from a quantum spin chain, since it does not involve non-universal
parameters like the sound velocity. However, the central charge alone does not
uniquely identify the CFT. The full operator content for a CFT on a ring of
length $L$ can be determined from the energy levels, which are spaced according
to the formula~\cite{DiFrancesco1997}
\be \label{cft_scal}
E_n - E_{GS} = \frac{ 2 \pi v }{ L } (  \Delta + m  + \overline{ \Delta } + \ell  )
\qquad m,\ell \in \mathbb{N}
\ee
where $n$ is a label for the $n$th excited state, $(\Delta,\overline{ \Delta
})$ are the weights of the two chiral representations of the Virasoro algebra
in the CFT, and $v$ is the non-universal sound velocity, which depends on the
microscopic realization of the CFT. The ground-state energy itself is affected
by universal finite-size corrections proportional to the central charge,
\be \label{gs_scal}
E_{GS} = \varepsilon_0 L  - \frac{  \pi v c }{ 6  L },
\ee
where $\varepsilon_0$ is the ground-state energy density in the thermodynamic
limit. Below, we analyze the spectrum obtained by exact diagonalization of the
lattice Hamiltonian, for systems with $L\le42$, in each momentum
sector~\footnote{Note that, in the $k=0$ sector, we have not split the spectra
according to their reflection symmetry, since the latter  seemed not to
correspond to charge conjugation symmetry of the two charged sectors}. After
extracting the central charge from the entropy scaling, Eq.~\eqref{gs_scal}
allows us to compute the sound velocity. The result we obtain by fitting the
ground-state energy for $L$ up to $42$ is $v=2.49(7)$. Another possibility is
to fit directly the dispersion relation of the low-energy states, which should
be linear and proportional to $v$. A sample of the low-lying spectrum is shown
in Fig.~\ref{velocity}(a) for a system of $L=39$. To obtain the velocity, we
perform a linear fit of the smallest available momentum at each system size.
The value of $v$ obtained in this way is different for right and left moving
particles, and in both cases deviates from the velocity extracted from the
ground state energy by a few percent [see Fig.~\ref{velocity}(b)]. This is
caused by large finite-size corrections affecting these eigenvalues. This
chiral symmetry breaking at finite size might be caused by the chiral
perturbation driving the system on the second-order transition line.
Interestingly, by taking the average of the corresponding right and left energy
levels, the dominant terms of these corrections cancel out, and full agreement
with the value extracted from the ground-state energy scaling is recovered.

Once the sound velocity is known, Eq.~\eqref{cft_scal} can be used to extract
all the conformal dimensions from the gaps in the low-energy spectrum of the
lattice Hamiltonian. The operators in the CFT are labeled by a $\mathbb{Z}_3$
quantum number~\footnote{The full symmetry group of the Potts model is the
permutation group $S_3 = \mathbb{Z}_3 \times \mathbb{Z}_2$. However, the
$\mathbb{Z}_2$ symmetry generator leaves invariant only $\mathbb{Z}_3$ neutral
states, namely the ones labeled by $Q=0$. Thus the $Q=0$ sector splits into two
subsectors labeled by a $\mathbb{Z}_2$ quantum number. Since this will not be
necessary to identify gaps with operator, it will not be used in the text.}
$Q=0,\pm 1$~\cite{Rittenberg1987,Rittenberg1986}. Since the model does not have
an exact $\mathbb{Z}_3$ symmetry, we have to find an alternative way of
labeling the low-lying states.

The $Q=\pm1$ sectors have to be degenerate and this degeneracy is exact at
finite size in the spectrum of the lattice Hamiltonian Eq.~\eqref{ham1} with
PBCs. This fact is ensured by the presence of the non-commuting momentum and
reflection symmetries, which implies that eigenstates of $H$ with momenta $K$
and $-K$ have the same energy. In the $\mathbb{Z}_3$-ordered phase and close to
it the states with momentum $K = \pm 2 \pi /3$ happen to be the lowest-energy
excitations above the ground state and the Brillouin zone appears to be
split in three, as shown in Fig.~\ref{velocity}(a). It is thus clear how to
identify the $\mathbb{Z}_3$ symmetry sectors: the neutral sector and the two
charged sectors consist of the energy levels close to $K=0$ and $K = \pm 2 \pi
/3$, respectively. This labeling naturally connects to the symmetry-breaking
structure of the ground-state manifold within the ordered phase.

The operator content of the two nondegenerate symmetry sectors in the CFT with
PBCs is~\cite{Rittenberg1987}
\begin{align}
\nonumber
& Q = 0 :  \quad  \( 0 , 0  \), \( \frac{2}{5}, \frac{2}{5}  \), \( \frac{7}{5}, \frac{2}{5}  \), \( \frac{2}{5}, \frac{7}{5}  \),  \\[1mm]
\label{op0}
& \hspace{2cm} \( \frac{7}{5}, \frac{7}{5}  \), \( 0 , 3  \), \( 3 , 0  \), \( 3 , 3  \)  \\[2mm]
\label{op1}
& Q = \pm 1  :   \hspace{1.4cm} \( \frac{1}{15}, \frac{1}{15}  \), \( \frac{2}{3}, \frac{2}{3}  \).
\end{align}
The eigenvalues of the lattice Hamiltonian are then spaced according to
Eq.~\eqref{cft_scal} and they correspond to the CFT operators above, with all
their descendants $(\Delta,\overline{\Delta})_{( k , \ell) }$. However, not all
the descendants are allowed and their degeneracy can be computed starting from
the Rocha-Caridi formula~\cite{Rittenberg1986}. The momentum of these states in
the CFT is instead given by
\be \label{cft_gaps}
\tilde{P} =  \frac{2 \pi}{ L } P = \frac{2 \pi}{ L } (  \Delta + m  - \overline{ \Delta } - \ell  )
\qquad m,\ell \in \mathbb{N}.
\ee

Note that the CFT momentum is not the lattice momentum for this Hamiltonian.
The CFT momentum on the lattice is measured starting from the ground states of
each $\mathbb{Z}_3$ sector, which we label by $P=0$ [see
Fig.~\ref{velocity}(a)].

We now proceed with the matching of the low-energy gaps on the lattice with the CFT prediction Eq.~\eqref{cft_scal}. Following Ref.~\onlinecite{Rittenberg1986}, we define the universal function
\be \label{un_f}
\mathcal{F}(Q,P) = \frac{L}{2 \pi v } \( E^{Q}_P - E_{GS} \)  \; \underset{L \to \infty }{ \sim }  \; \Delta + k + \overline{\Delta} + \ell ,
\ee
where $Q$ and $P$ are the CFT $\mathbb{Z}_3$ quantum number and momentum.

\begin{figure}
{\includegraphics[scale=0.5]{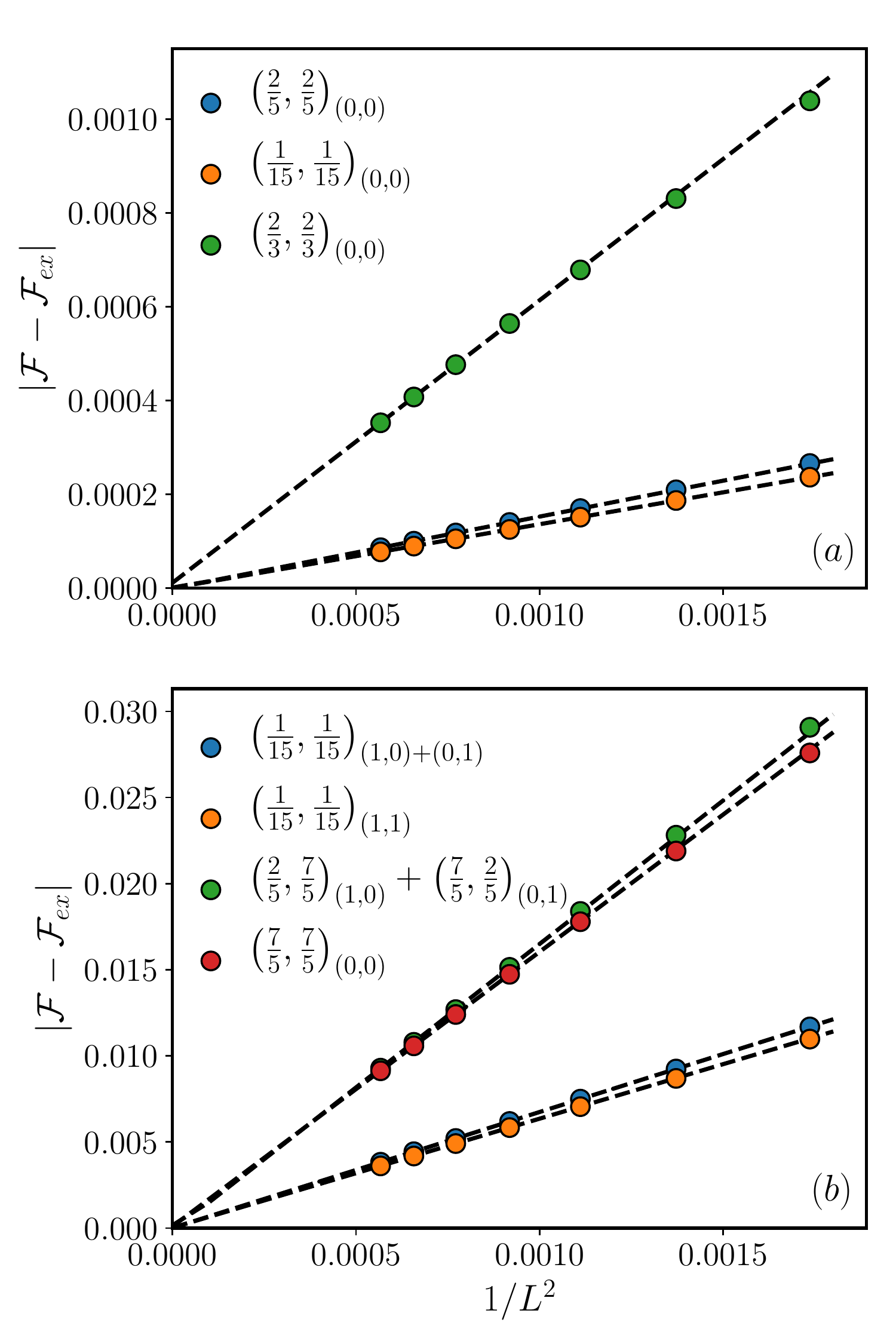}\hspace{2mm}}
\caption{Finite-size scaling of the universal function $\mathcal{F}$ in
  Eq.~\eqref{un_f} with respect to the CFT expected value. (a) First and second gaps in
  the $Q=1$ sector (orange and green) and first gap in the $Q=0$ sector (blue).
  Finite-size corrections scale as $L^{-2}$ with a coefficient of magnitude
  $10^{-4}$ for the first two gaps. (b) Third gap in $Q=1$ sector (orange),
  first gap in the $Q=1$ sector with momentum $P=\pm1$ (blue), average of the
  second of fifth gap (green) and third gap (red) in the $Q=0$ sector. The
  finite-size corrections are always quadratic in the inverse length of the
  chain upon appropriate average between CFT states not invariant under
  $(\Delta,k) \leftrightarrow (\overline{\Delta},\ell)$.}
\label{gap1}
\end{figure}

The results of the field correspondence are presented in Fig.~\ref{gap1}. Upon taking proper combinations of degenerate gaps, the finite-size corrections are of order $L^{-2}$ for all the gaps, with a prefactor smaller than $10^{-3}$ for the lowest ones. We extrapolate the value of $\mathcal{F}$ by a two-parameter fit for system sizes up to $L=42$. The agreement of the extrapolation with the CFT expected values is perfect once the sound velocity is tuned to $v = 2.49225$. In this respect, this method is the best way to estimate the sound velocity with the available system sizes.

The finite-size corrections to the universal function in Eq.~\eqref{un_f} have been studied for this universality class in the 3-states Potts chain~\cite{Rittenberg1987}. It was observed that their power-law exponent was $2$ for most of the nondegenerate gaps and a number between $0.5$ and $1$ for other degenerate gaps. Here we argue that the latter corrections appear only in CFT states $(\Delta,\overline{\Delta})_{(k,\ell)}$ for which $(\Delta,k) \ne (\bar{\Delta},\ell)$. Upon taking the average of the eigenvalues in which $\Delta,k$ and $\overline{\Delta},\ell$ are exchanged these dominant corrections vanish. Formally
\begin{align}
\mathcal{F}( \Delta, \overline{\Delta} , k , \ell ) = \Delta + \overline{\Delta} +  k +  \ell + \frac{ A^{(\Delta, \overline{\Delta})}_{k,\ell}   }{ L^\gamma } +  \frac{ B^{(\Delta, \overline{\Delta})}_{k,\ell} }{ L^2 } + \dots \nonumber\\
A^{(\Delta, \overline{\Delta})}_{k,\ell}  = - A^{(\overline{\Delta}, \Delta)}_{\ell,k}  \hspace{2.5cm}
\end{align}
To support this statement we give two examples where this is manifest. We take
the lowest-lying pair of states in the $Q=1$ sector of $H$ with momentum $P =
\pm 1$, i.e., $(1/15,1/15)_{(1,0)}$ and $(1/15,1/15)_{(0,1)}$. We then take the
pair of states in the $Q=0$ sector with momentum $P=0$, i.e.,
$(2/5,7/5)_{(1,0)}$ and $(7/5,2/5)_{(1,0)}$. On the spin chain these two states
correspond to the second and fifth excited state in the $K=0$ sector. Their
finite-size scaling is plotted in Fig.~\ref{gap1}(b), where these two gaps are
denoted by blue and green circles, respectively. The same agreement is observed
with many other levels not reported here. We are able to match irrelevant CFT
operators with large conformal weights as $(0,3)$, $(3,0)$, and $(3,3)$ and the
rule for which the dominant finite-size corrections cancel still applies.

\subsection{Density and order parameter two-point functions}

It is, in general, a difficult task to associate matrices on the lattice to
primary fields in the CFT. The operator for which this procedure is trivial is
the order parameter, namely the most relevant operator in the CFT which is not
invariant under a symmetry transformation, i.e., the primary field
$(1/15,1/15)$. Its anomalous dimension is thus $\eta_\sigma = 4/15$ and its
two-point function is expected to behave as a power law with this exponent.
$\mathbb{Z}_3$ order is realized on the lattice through a period-3
boson-density wave; thus the (complex) order parameter takes the
form~\cite{Fendley2004}
\be \label{lat_op}
O_i = n_i + e^{ i 2 \pi / 3 } n_{i+1} + e^{ - i 2 \pi / 3 } n_{i+2}.
\ee
Exploiting translational invariance, we can write its two-point function in
terms of the density two-point function as
\begin{align}
\nonumber
\langle O^\dagger_{\mathstrut r } O_{\mathstrut 0  }   \rangle  & = 3 \langle n_0 n_{r}  \rangle + e^{ i 2 \pi / 3 } \big(  2 \langle n_0 n_{r+1}  \rangle +  \langle n_0 n_{r-2}  \rangle  \big)  +  \\[1mm]
& \quad  + e^{- i 2 \pi / 3 } \big(  2 \langle n_0 n_{r-1}  \rangle  + \langle n_0 n_{r+2}  \rangle  \big) .
 \label{lat_2pf}
\end{align}
\noindent
If translational invariance can be assumed in the system (as in our case), this
quantity will be purely real; a very small imaginary part will be obtained when
determining $\langle O_r^{\dagger} O_0 \rangle$ from numerical data, and will
be neglected.
In order to take into account finite-size effects, we compare our results to
the two-point function of the order parameter for a CFT on a ring of length
$L$. For a primary field with conformal weights $\Delta = \overline{\Delta} =
\eta/4 $, this quantity reads~\cite{DiFrancesco1997}
\be \label{cft_2pf}
\langle O(x) O(y) \rangle = \frac{A}{ \[ L \sin \( \frac{ \pi (x-y)}{L} \)
\]^\eta  } =  \frac{A}{ L^\eta } G\(  \frac{x-y}{ L } \) .
\ee

\begin{figure}
{\includegraphics[scale=0.5]{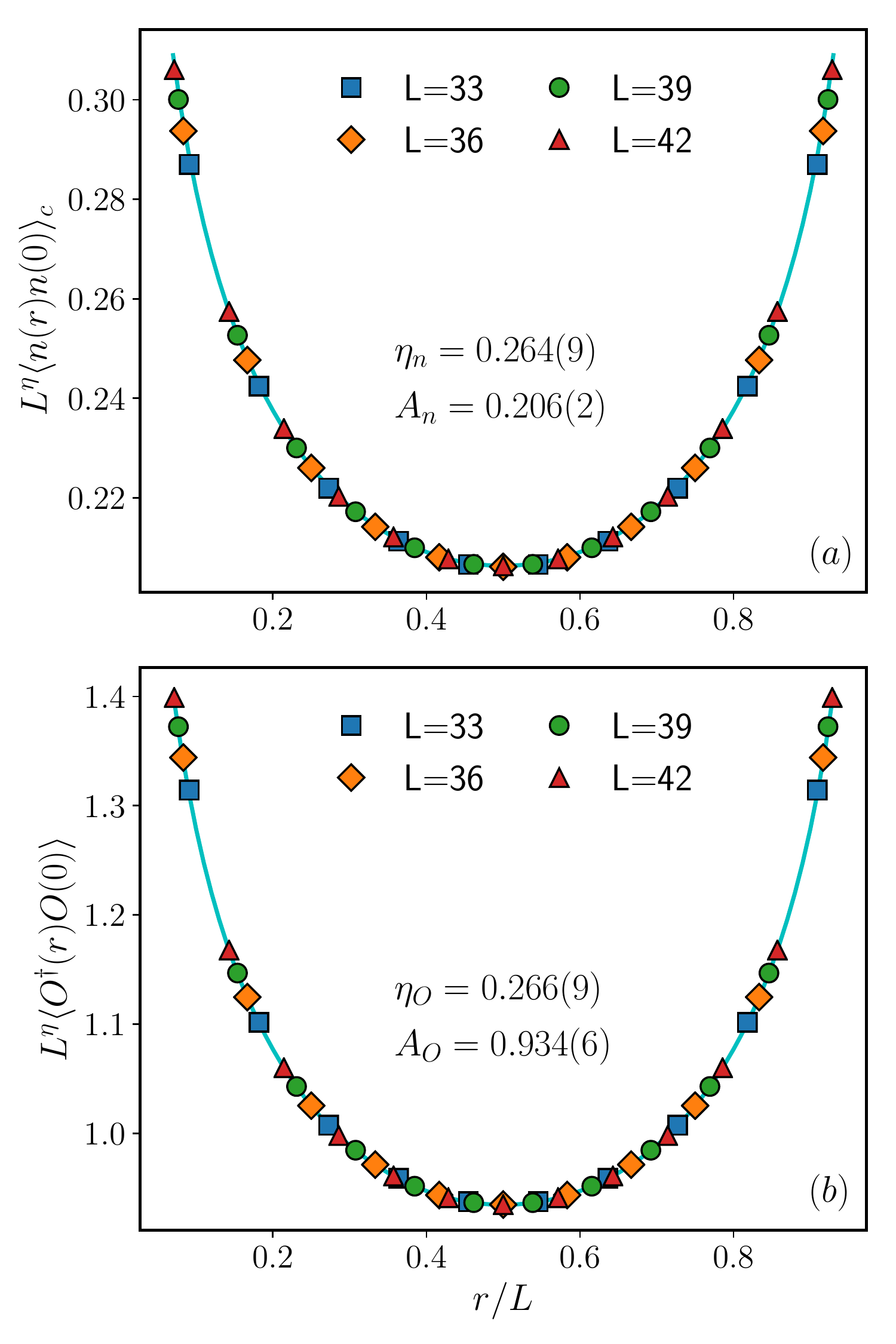}\hspace{2mm}}
\caption{(a) Two-point function of the lattice order parameter
  Eq.~\ref{lat_2pf} for different lengths $L$ multiplied by $L^\eta$, with
  $\eta$ fitted with the CFT expression Eq.~\ref{cft_2pf}. Estimate and error
  of the amplitude $A$ and the exponent $\eta$ are obtained upon taking the
  average of the results for different system sizes. (b) Same scaling as in (a)
  for the order parameter, which has the same scaling dimension as the
  density.}
\label{twopf}
\end{figure}

We can then obtain an estimate of $\eta$ by fitting the lattice two-point
function with the expression above and free parameters $A$ and $\eta$. In
Fig.~\ref{twopf}(b) we plot the lattice expectation value for different system
sizes, rescaled by multiplication by $L^{\eta}$ (where the value resulting from
the fit mentioned above is taken for the latter), obtaining perfect data
collapse on the universal scaling function $G(x)$ in Eq.~\ref{cft_2pf}. In
Fig.~\ref{twopf}(a) we plot the connected density-density expectation value,
which also fits perfectly the CFT expression, with the same scaling dimension
as the order parameter.

\section{Doubly-blockaded regime}
\label{sec4}

In this section, we study the phase diagram of the model Hamiltonian in
Eq.~\eqref{ham1} in the limit $V \to +\infty$.  When $U \to -\infty$ the system
is $\mathbb{Z}_3$-ordered and the order parameter in Eq.~\eqref{lat_op} is non
vanishing. For finite and large negative $U$, the finite-size spectrum of the
Hamiltonian behaves as in the usual $\mathbb{Z}_3$ spontaneously
symmetry broken scenario: the ground state is nondegenerate and the
first two low-lying excited states are exponentially close to it with a gap
$\Delta \propto \exp( -L/\xi )$, where $\xi$ is the correlation length.  In the
limit $U \to +\infty$ the ground state is the nondegenerate state with no
bosons and $\mathbb{Z}_3$ symmetry is not broken. A transition between these
two regimes is expected in the middle.

In what follows, we provide evidence that there are two continuous phase
transitions located at $U_{c 1} \lesssim -1.96$ and $U_{c 2} \simeq -1.915$.
At the first transition, the ground state of the system switches from a
period-3 ordered state to a quasi-long-range-ordered, critical phase with
incommensurate density-density correlations, known as the \textit{floating phase}.
At the second transition point, the system passes from the gapless critical
phase to a disordered phase. For the first transition we compute, with
different methods, the location of the critical point, the correlation length
critical exponent $\nu$, the dynamical critical exponent $z$, and the order
parameter critical exponent $\beta$. We then show that the second transition is
consistent with the BKT scaling ansatz, according to which the correlation
length vanishes exponentially and the gap finite-size scaling at the transition
point is affected by logarithmic corrections~\cite{BKTGAP1987}.

We finally show that for values of $U$ inside the floating phase $U_{c 1} \le U
\le U_{c 2}$ the scaling of the entanglement entropy is in agreement with the
Luttinger liquid universality class, where the central charge $c$ equals 1.

\subsection{Quantum concurrence and fidelity susceptibility}

By means of the same methods tested in Sec.~\ref{concfid_potts}, we now proceed
to investigate the transition points by studying the behavior of the quantum
concurrence and the fidelity susceptibility. These observables are not known to
be generically sensitive to BKT transitions~\cite{Sun:2015aa}; for this reason,
we expect them to only diagnose the presence of the first of the two
transitions mentioned above. We carry out exact diagonalization calculations up
to $L = 54$ sites for ground-state properties, and consider sizes $L = 3n, n
\in \mathbb{N}$ to avoid incommensurability effects.

\begin{figure}
{\includegraphics[scale=0.5]{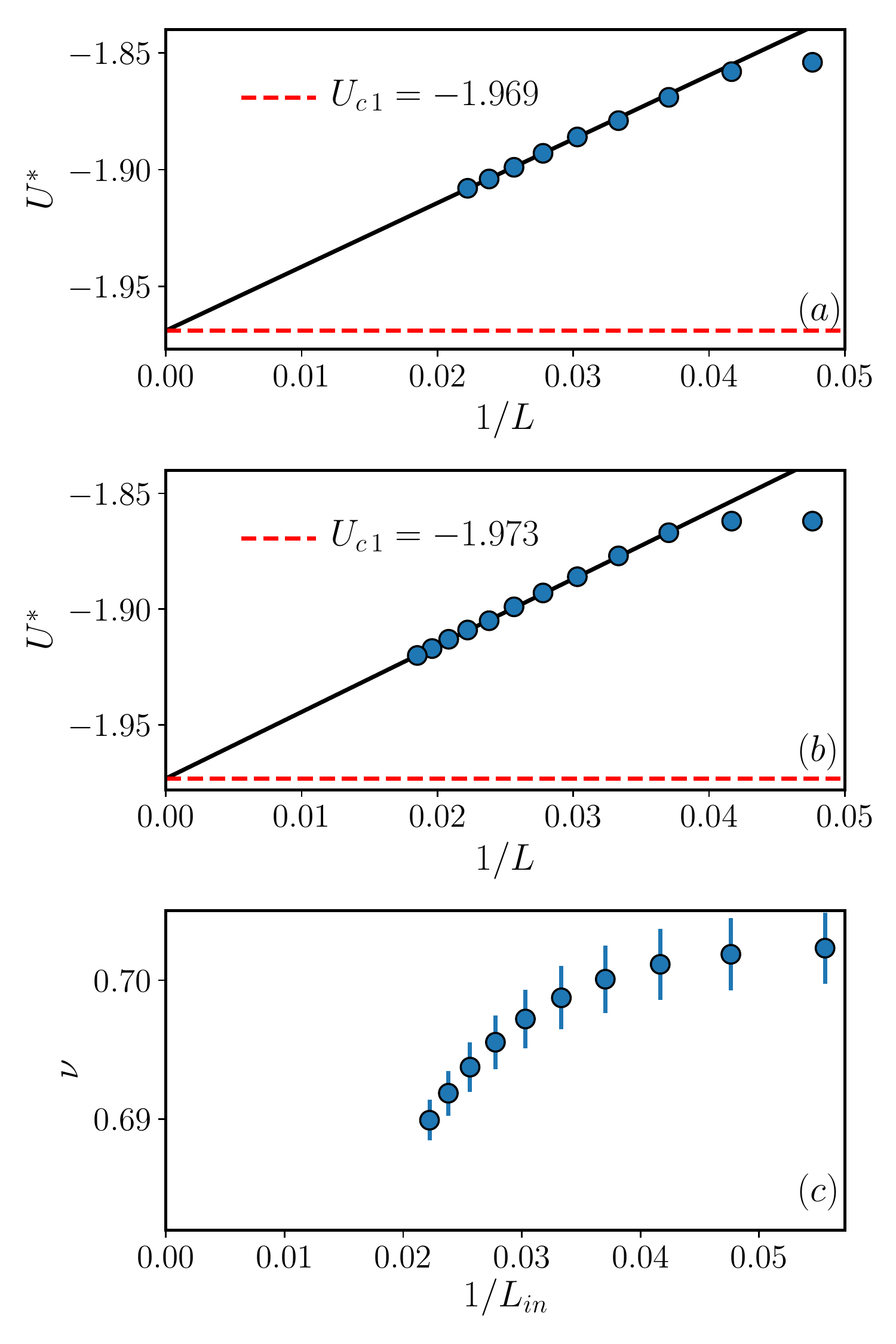}\hspace{2mm}}
\caption{(a) Linear fit of the peak position $U^*(L)$ of the first derivative
  of the quantum concurrence Eq.~\eqref{conc} vs $1/L$, for $L$ from 33 to 45
  sites. The result of the fit is stable against the system sizes included in
  the fit and the critical position we obtain $U_{c1} = -1.969 \pm 0.005$,
  where the error takes into account variations with respect to the system sizes
  included in the fit. (b) Linear extrapolation of the peak position $U^*(L)$
  vs $1/L$ in the fidelity susceptibility Eq.~\eqref{fid_susc}, for $L$ from 39
  to 54 sites. The result is stable when smaller system sizes are included and
  the critical point position we get is $U_{c 1} = -1.973 \pm 0.005$. Error
  considerations are the same as in panel (a). (c) Correlation length critical
  exponent obtained from the scaling of the maximum of $\chi_F$ according to
  Eq.~\eqref{fid_scal} for $L=L_{in},L_{in}+3,...,54$ as a function of
  $L_{in}$. The critical exponent decreases when smaller system sizes are
  excluded from the fit and saturation is not reached with the maximum lengths
  we can access. We note that a strong sensitivity of critical exponents with
  respect to system sizes was already noted in Ref.~\onlinecite{Chepiga2018}.}
\label{fid_conc_inf}
\end{figure}

The derivative of the concurrence exhibits the same behavior discussed in
Sec.~\ref{concfid_potts}, namely a peak which is sharpening and moving towards
the critical point, $U_c$, with increasing system size. In order to extrapolate
the position of the maximum $U^*(L)$ for $L\to \infty$ we fit it with a
power law with scaling exponent $\gamma = 1$. In this way, we obtain a value
$U_{c 1}  = -1.969 \pm 0.005$, which is stable against the range of system
sizes included in the fit. By performing the same analysis for the peak of the
fidelity susceptibility, Eq.~\ref{fid_susc}, we get instead a critical value
$U_{c 1} = -1.973 \pm 0.005$. The results are plotted in
Fig.~\ref{fid_conc_inf}(a)-(b). Both of these results illustrate the fact that
finite-size effects in this regime are comparatively larger than close to the
Potts critical point. In particular, employing sizes on the order of $L\simeq 30$
would lead to wrong estimates in both cases: the scaling regime for what
concerns entanglement and wave-function properties seems to be only reached
above $L= 33$.

Exploiting the known finite-size critical scaling of the peak of $\chi_F$
described by Eq.~\ref{fid_scal}, we obtain a critical exponent $\nu = 0.70 \pm
0.05$. We stress that the latter estimate is very sensitive to the system sizes
employed in the analysis. In particular, the larger the system sizes included
in the fit, the smaller the $\nu$ obtained [see Fig.~\ref{fid_conc_inf}(c)]. In
Ref.~\cite{Samajdar2018}, the value $\nu \simeq 5/7$ was extracted from data
collapse of the gaps, assuming the value $z\simeq 4/3$ for the dynamical
critical exponent. This evidence was used to conclude that the transition does
not belong to the PT universality class, for which $\nu = 1/2$ and $z=2$.
Although the variation of the exponent with the system sizes considered seems
very slow, we cannot exclude, based on our data, that it eventually reaches the
value expected for a phase transition of the PT type, as found in
Ref.~\onlinecite{Chepiga2018} for the critical regime below the Potts point. 

\begin{figure*}[t]
\hspace{-9mm}{\includegraphics[scale=0.47]{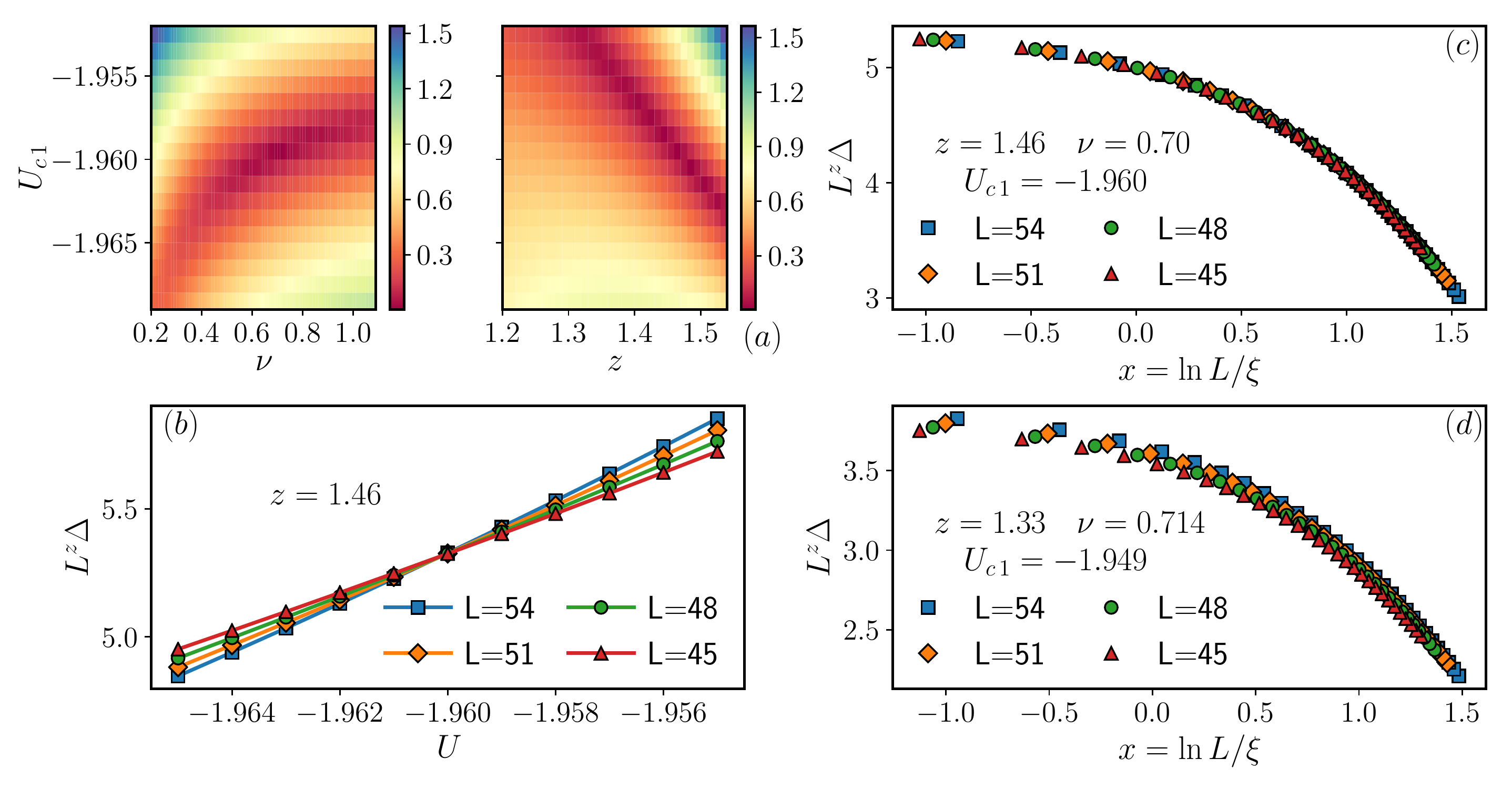}\hspace{-2mm}\includegraphics[scale=0.47]{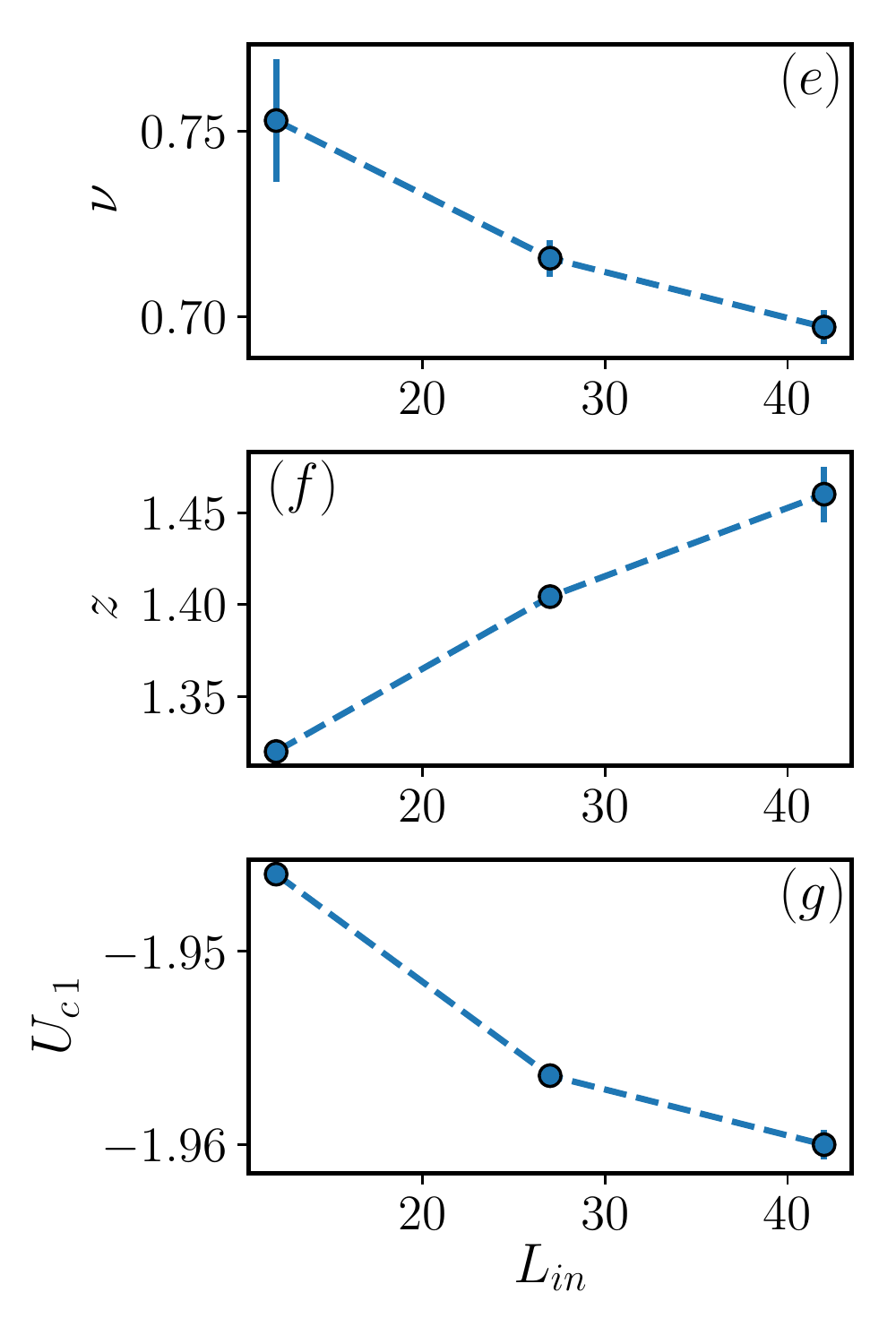}}
\caption{(a) Density plot of the square root of the sum of the squared
  residuals in the $(\nu,U_{c1})$ and $(z,U_{c 1})$ planes for the best-fitting
  values of $z$ and $\nu$, respectively. (b) Crossing of the gaps, upon
  multiplication by $L^z$ for the best fitting $z$. The crossing indicates the
  position of the critical point. (c) Data collapse of ED numerical data, with
  $U \in [U_{c 1} - 0.03 ,U_{c  1}]$ with the parameters $U_{c 1}, z,\nu$ which
  minimize the polynomial fit of the universal scaling function in
  Eq.~\eqref{scal_znu}. (d) Same as in (c), but with parameters $U_{c 1},
  z,\nu$ taken from Ref.\onlinecite{Samajdar2018}. (e)-(g) Critical exponents
  and critical point location obtained by applying the procedure described in
  Sec.~\ref{procedure} for sets of 5 system sizes $L = L_{in} , ... , L_{in} +
  12$ with increasing $L_{in}$. The average is obtained by varying the size of
  the interval $[U_{c 1} - \delta U , U_{c 1} ]$ with $0.01\le  \delta U \le
  0.03$, the degree of the polynomial being fixed to 10. The errorbar is the
  standard deviation of the obtained results.}
\label{gap_znu}
\end{figure*}

\subsection{Critical point location through data collapse}
\label{procedure}

We now exploit the finite-size scaling theory which applies in the proximity of
a second-order phase transition\cite{Fisher1989} to extract the values of $U_{c
1}$, $z$, and $\nu$ from the lowest spectral gap $\Delta$. With this aim, we
adopt an approach very similar to the one taken in Refs.~\onlinecite{Rigol2011,
Dalmonte2015}.

First we compute a universal scaling function $F$ from ED data for $\Delta$ for
different system sizes. This scaling function will depend on some unknown
critical exponent $\gamma$ and on the dimensionless ratio $L/\xi$, $L$ being
the system size and $\xi$ the correlation length: $F = F(\gamma,L/\xi)$. We
then assume a functional form for $\xi$ in terms of the critical point position
$U_c$ and, if finite, of its critical exponent $\nu$: $\xi = \xi(U,U_c,\nu)$.
We use this functional form to express the scaling function $F$ in terms of the
variable $x = \ln( L/\xi )= x(L,U,U_c,\gamma,\nu)$. Finally, we combine data
for $F$ for different system sizes and we look for the values of
$U_c,\nu,\gamma$ which produce the best data collapse. This is achieved by
fitting $f(x) = F( e^x )$ with an arbitrary high-degree polynomial and
minimizing the sum of the squared residuals. By considering a full functional
collapse instead of extracting the thermodynamic limit gap from single
parameter data, this method copes relatively well with finite-size effects,
even in the most critical BKT scenario. Indeed, in the latter case it allows us to
locate transition points with a precision similar to (if not better than) approaches
based on matching conformal dimensions~\cite{Rigol2011, Dalmonte2015}, which are
based on assuming a specific functional dependence between lattice and field
theory operators.

Since at a quantum phase transition all low-lying eigenvalues of the
Hamiltonian are expected to be separated from the ground state by a power-law
decaying gap $\Delta \sim 1/L^z$, where $z$ is the dynamical critical exponent,
we can obtain a scaling function by multiplying the lowest gap by $L^z$:
\be
\label{scal_znu}
 F \(  \frac{L}{\xi}  \) = L^z \Delta \,.
\ee
Assuming that the phase transition has a finite $\nu$ exponent, we have
\be
\label{scal_xi}
\xi \sim \( U_{c 1 } - U \)^{-\nu } .
\ee
We can then find the best-fitting values of $\nu$,$z$, and $U_{c 1}$ via the
procedure described above. It is fundamental to check the stability of the
result with respect to the degree of the polynomial, the size of the interval from which
the value of $U < U_{c 1}$ is taken, and most importantly, the system sizes
which are included in the fit. We find that, in our case, the result is very
stable with respect to the first two, but we get stability with respect to the system sizes
we have at our disposal only if we include the largest ones (up to $L_{\max} =
54$). In particular, by including sizes of increasing magnitude we observe a
decrease in our estimates for $U_{c1}$ and $\nu$, and an increase in the one of
$z$. In Fig.~\ref{gap_znu}(c) we show the result obtained by including all
systems sizes $45 \leq L \leq 54$. The data collapse shows deviations of order $10^{-2}$, and is considerably more accurate
than the one performed with the values reported~\cite{Samajdar2018} with
$L_{\text{max}}=36$ [see Fig.~\ref{gap_znu}(d)]. 

However, as shown in Figs.~\ref{gap_znu}(e)-(f), the $z$ and $\nu$ exponents are
still varying with the system size. Although the trend exhibited by this data
does not allow any extrapolation, we clearly see that the true scaling regime
has not yet been reached. This leaves open the possibility that $z$ and $\nu$
will eventually approach the values expected from a JNPT transition, namely $2$
and $1/2$, respectively~\cite{PT1979}. The best estimates we can give from our
data of critical exponents and critical point position are: $z = 1.48 \pm 0.1$,
$\nu = 0.7 \pm 0.1$, $U_{c 1} = -1.960 \pm 0.005$. 

Since the methods employed over the next subsection will rely on assumptions,
we find it useful to summarize the analysis performed so far. All diagnostics
are compatible with the presence of a second order phase transition. The
location and nature of the transition are extremely sensitive to the system
sizes investigated. Regarding the location of the transition point, sizes up to
$L\simeq 30$ are not sufficient to determine it, while the estimates using all
three methods are rather stable after $L\simeq 45$. Entanglement-based methods
return $U_{c1}=-1.973\pm 0.005$ and $U_{c1}=-1.969\pm 0.005$, respectively. The
method based on gap collapse returns $U_{c1}=-1.960$; for this last method, it
is challenging to include a rigorous error bar. However, it is worth noting
that the best data collapse obtained up to $L=36$ returns $U_{c1}=-1.949$, in
agreement with Ref.~\onlinecite{Samajdar2018}; this clearly signals that the
critical point is drifting to considerably smaller values of $U$ as size
increases [see Fig.~\ref{gap_znu}(g)], in agreement with the entanglement-based
diagnostics. 

A similar conclusion holds for the critical exponents: as clearly observed in
the fidelity susceptibility scaling, even at sizes of order $L=54$, the
critical exponent has not yet reached its thermodynamic value. The data
collapse of the finite-size gaps fully confirms this picture. This motivates
the study in the next subsection, where we will employ different -- but
assumption-dependent -- methods to determine some of the properties of this
second-order transition. From the analysis performed here, we can anticipate
that, even if larger system sizes are studied, depending on the observable, a
systematic underestimation of the modulus of the critical point location
$|U_{c1}|$ is expected. As we will see, this is particularly critical
for the methods discussed in the next section.

\subsection{Order parameter}

In this section, we investigate the disappearance of the $\mathbb{Z}_3$ order
across the second-order phase transition. We do so by utilizing three methods:
a QMC and DMRG study in PBCs, and a 1-site DMRG study with OBCs. Our focus in the
following will be on correlation functions and the order parameter of the
$\mathbb{Z}_3$ order. As such, we will be assuming that there is an exact
mapping between the lattice operators describing the latter, and its field
theory counterpart. While this condition is typically satisfyingly fulfilled
for most lattice models displaying quantum critical behavior, we opted for a
more conservative approach in the context of the FSS model in the doubly blockaded
regime. The reason is that the constraint acts at the lattice spacing level
irrespectively of how close one is to the critical point. This suggests that
defining field operators that do not change at the lattice spacing level are
nontrivial, making the connection between lattice and continuum not immediate.
While this feature has no consequence on spectral and wave-function properties,
it is highly likely that it affects the finite-size behavior of correlation
functions.

We perform quantum Monte Carlo (QMC) simulations using a modified version of
the worm algorithm~\cite{Prokofev1998-1}, adapted to simulate Hamiltonians with
off-diagonal terms such as those of the FSS model and with updates designed to
automatically respect its occupation constraints. The method allows us to directly
measure quantities such as energy, particle density, and the static structure
factor:
\begin{equation}
  S(k) \equiv \frac{1}{L^2} \sum_{i, j = 1}^L e^{-i k  
 r_{ij}} \ave{n_i n_j},
\end{equation}
where $k$ is one of the allowed lattice momenta, and $r_{ij}$ is the distance
between sites $i$ and $j$. In the $\mathbb{Z}_3$--ordered phase the structure
factor will feature a peak at a wave vector $k = 2\pi/3$, corresponding to the
periodicity of the $\mathbb{Z}_3$-periodic charge density waves. The value of
the peak is equal to the squared $\mathbb{Z}_3$ order parameter in
Eq.~\eqref{lat_op}, and therefore follows a power-law behavior
\cite{Wang2006} $S(2\pi/3) \sim |U - U_c|^{2 \beta}$ when approaching the
critical point $U_c$ from the ordered phase.

\begin{figure}
  \includegraphics[width=\columnwidth]{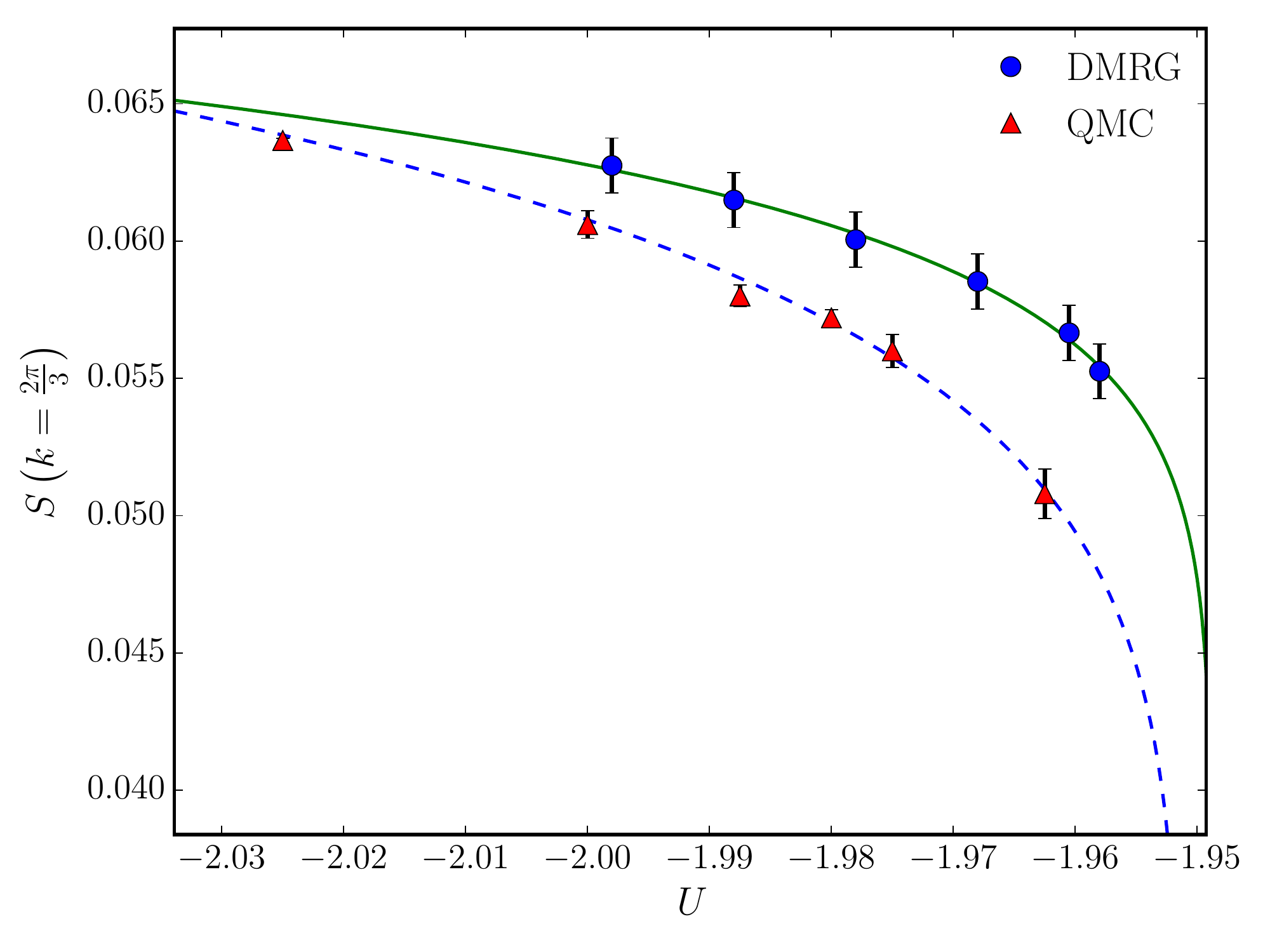}
  \caption{$S(2 \pi/3)$ as a function of the parameter $U$. Symbols
  represent QMC (triangles) and DMRG (circles) for the ground state of model
  Eq.~\eqref{ham1}, extrapolated in the thermodynamic limit (see text). The
  dashed and solid line are a power-law fitting function (see text) used to
  interpolate QMC and DMRG results, respectively.}
  \label{fig.mc_dmrg} 
\end{figure}

We obtained an estimation of the position of the critical point, as well as the
critical exponent $\beta$, by interpolating the QMC results with the expected
power-law behavior. We studied system sizes up to $L = 120$ sites and
temperatures down to $T = 1/128$ (where the magnitude of the off-diagonal part
of the Hamiltonian is taken as a unit of energy). Extrapolation in the inverse
temperature has been employed to determine ground-state results where direct
convergence in $T$ [i.e., results identical within their uncertainty for one or
more pairs of temperatures $(T, T/2)$] was not observed. Below $U = -1.96$, no
further extrapolation in the system size was necessary, since direct
convergence in size was always observed. Above this value, however, our
extrapolated values were not fully converged in size and inverse temperature
(also due to considerably slower MC dynamics). We therefore restricted our
investigation to the $U < -1.96$ region. Figure~\ref{fig.mc_dmrg} shows the
QMC data (triangles) as well as the power-law interpolation (purple line). The
resulting values are $U_{c 1} = -1.951(5)$, $\beta = 0.059(7)$~\footnote{These
confidence intervals only indicate the error in the fit; the same applies to
the DMRG estimate of the same quantities.}.

An independent estimate of the critical point and critical exponent $\beta$ has
been obtained by computing $S(2 \pi/3)$ via DMRG (circles in
Fig.~\ref{fig.mc_dmrg}) and performing the same extrapolation as above (solid
green line in Fig.~\ref{fig.mc_dmrg}). We approximated the exact Hilbert space
in the DMRG by giving a large penalty to not-allowed states. This is achieved
by adding to the Hamiltonian a term $\lambda \sum_i n_i n_{i+1} + n_i n_{i+2}$,
with $\lambda=10^3$. Unfortunately, performing a rigorous extrapolation in
$\lambda\rightarrow\infty$ is extremely difficult: the main reason is that, for
increasingly larger values of $\lambda$, the diagonalization at each DMRG step
becomes extremely sensitive to numerical errors due to the large difference in
the matrix elements of the Hamiltonian matrix. However, for a fixed value of
$\lambda$, we expect a difference on the order of $1/\lambda$ when comparing
local observables, such as energies, with ED data. The absence of any other
unforeseen source of systematic error due to the finite value of $\lambda$ can
be confirmed by direct verification. Indeed, with our choice of $\lambda=10^3$,
if we calculate the energy gap between the ground state and the first excited
level for $L=54$ and $U=-1.950$, the discrepancy between DMRG and ED is of the
(expected) order of $\epsilon_{\Delta} \approx 10^{-3}$. This check is very
important because it allows us to understand that the limit $\lambda
\rightarrow \infty$ is approached perturbatively. Despite this violation
of the constraint which directly affects local observables, we obtain a good
agreement with ED when we study other quantities such as entanglement entropy
and central charge. For instance, using the same values of $L$ and $U$, we
obtain a difference in the central charge on the order of $\epsilon_{c} \approx
10^{-4}$ with respect to ED results. In our DMRG implementation, we take an
elementary cell made of $3$ sites in order to have a local representation of
the $\mathbb{Z}_{3}$ order. This also allows us to discard $4$ of the $8$
states in the blocked DMRG-site. Simulations were performed by keeping the
truncation error below $10^{-7}$ using up to $1000$ DMRG states and ensuring
that the energy variance of the ground state is of the same order of the
truncation error.

We observe that DMRG results (after an extrapolation in $1/L$ of the squared
$\mathbb{Z}_{3}$ order parameter for $84 \leq L \leq 120$) yields a larger
value for $S(2\pi/3)$ than QMC, possibly due to the approximations required to
impose the occupation constraint. The results of the extrapolation are $U_{c 1}
= -1.948 \pm 0.007$ and $\beta = 0.036 \pm 0.005$. 

\begin{figure}
{\includegraphics[scale=0.5]{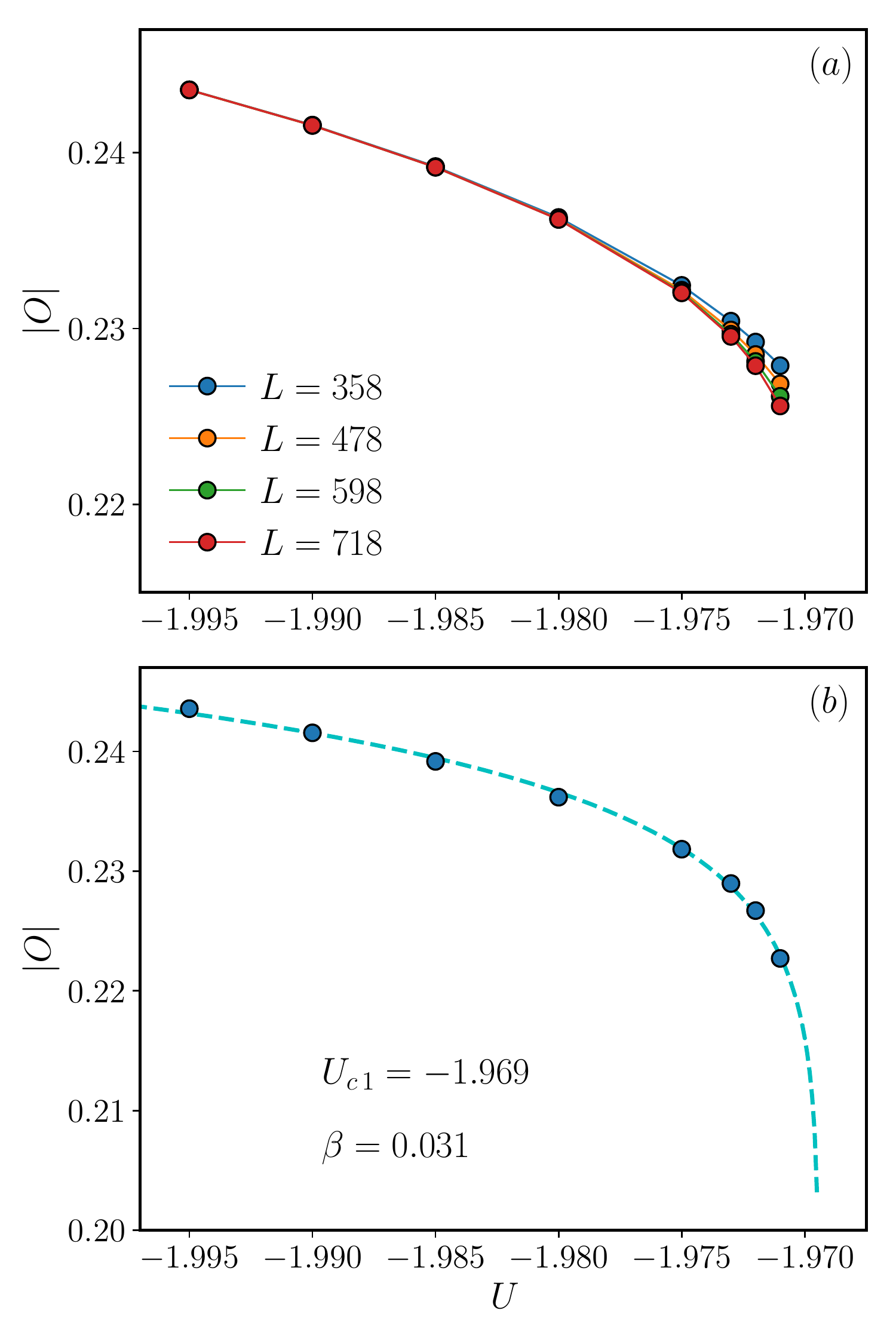}\hspace{2mm}}
\caption{(a) Order parameter computed by averaging the one-point function $O_j
  = e^{i 2 \pi j/3} n_j$ on $L/2$ sites in the bulk. $\mathbb{Z}_3$ symmetry is
  spontaneously broken by the choice of the number of sites on the open chain,
  i.e., a multiple of 3 plus 1 site. The obtained order parameter does not
  scale with the system size for $U < -1.975$. (b) Infinite-size limit value of
  the order parameter, extrapolated in $1/L$ and power-law fit of the resulting
  curve. The obtained critical exponent and critical point position are $\beta
  = 0.031 \pm 0.005$ and $U_{c 1} = -1.969 \pm 0.002$.}
\label{mps_op}
\end{figure}

As a final test bed for the results above, we compute directly the order
parameter in Eq.~\eqref{lat_op} by variationally optimizing the ground state
with MPS methods on an open chain in which the constraint is implemented
exactly. The method we use exploits the exact relation between MPOs and
finite-state automata~\cite{Crosswhite2008}, and is described in detail in the
Appendix. We are able to variationally optimize the MPS for chains of up to
$718$ sites. The computational resources required to accurately approximate the
ground state are relatively small: a bond dimension of $200-300$ is sufficient
to keep the variance of the Hamiltonian below $10^{-9}$. We explicitly break
$\mathbb{Z}_3$ symmetry by choosing system sizes which are multiples of 3 plus 1
site. This makes energetically favored states in which there are two bosons at
the edges, thus breaking the symmetry without adding any term in the
Hamiltonian. Extrapolation to the thermodynamic limit is then performed vs
$1/L$. The result is plotted in Fig.~\ref{mps_op}. We compute the order
parameter by averaging the one-point function on $L/2$ sites in the bulk. The
fit of the averaged order parameter as a function of $U$ returns a critical
exponent $\beta = 0.031 \pm 0.005$ and a critical point location $U_{c 1} =
-1.969 \pm 0.002$. The error attributed takes into account variations of the
fitting parameters obtained by considering different sets of values of $U$ and
computing the order parameter by performing the average over a different number
of sites in the bulk of the chain. 

Summarizing, the direct study of the order parameter provides similar
information to that of the quantities analyzed in the previous subsection: upon
increasing system sizes, the position of the second-order transition
systematically drifts toward larger values of $|U_{c 1}|$. It is informative to
note that this shift is compatible with a ``finite-size'' location of the
transition point based on the wave-function variation captured by the fidelity
susceptibility: as can be seen from Fig.~\ref{fid_conc_inf}(b), a
finite-size estimate at around $L\simeq 120 / 800$ would return a critical
coupling of order $U_{c1}\simeq -1.95 / 1.97$, respectively. The
incompatibility with the extrapolated values of the structure factor between
DMRG and QMC indicates that approaching an exactly blockaded regime in
experiments is challenging (see, e.g., the relatively large deviations in
estimating $\beta$), even if, in terms of transition point location, the
difference is of order $0.003$.

\subsection{BKT transition and the floating phase}

\begin{figure}
{\includegraphics[width=\columnwidth]{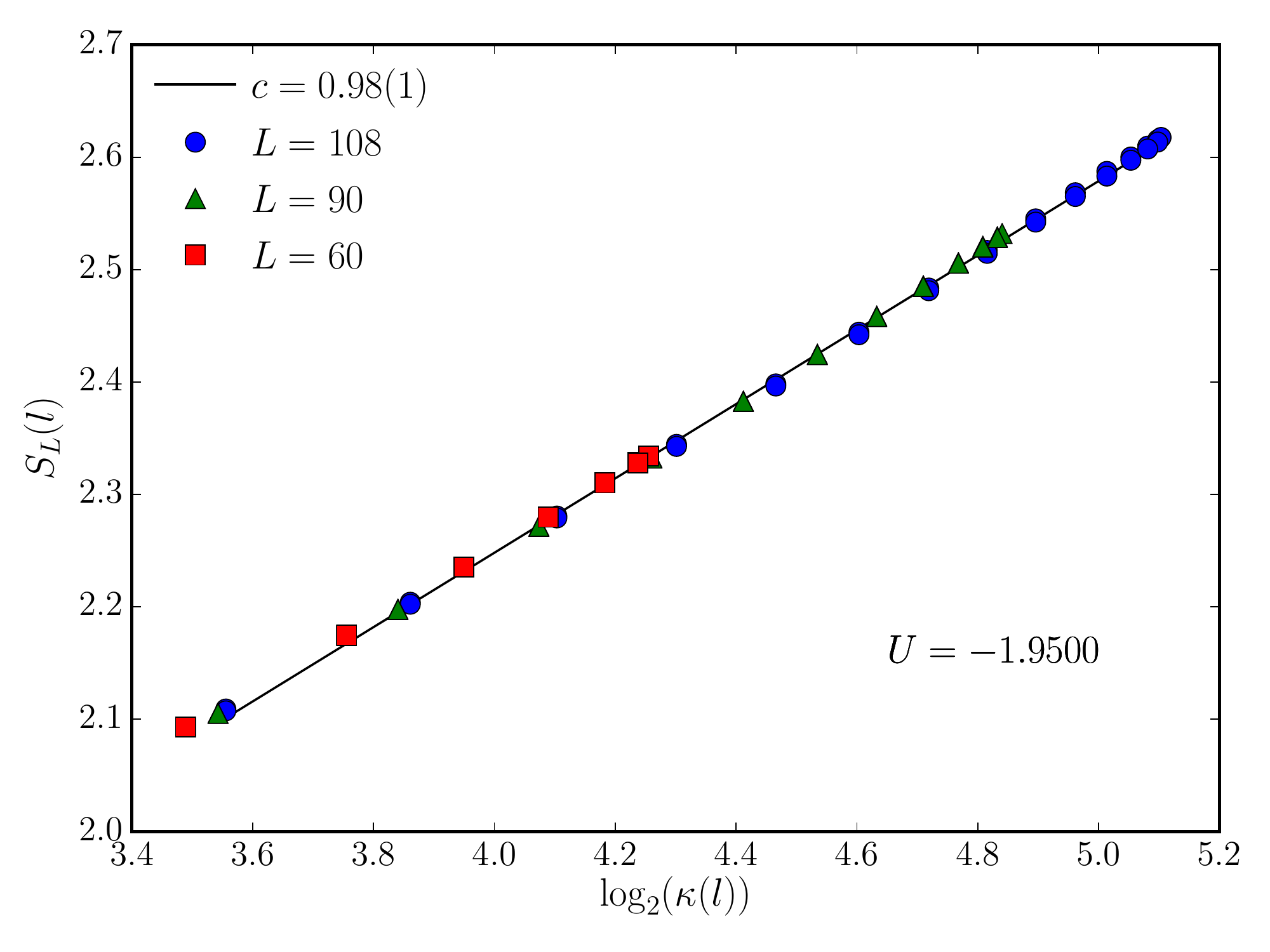}\hspace{2mm}}
\caption{Entanglement entropy for fixed $U$ inside the floating phase as a
  function of the logarithm of the cord length in the CFT ring. The fit
  produces a central charge in a perfect agreement with the Luttinger liquid CFT.
  }
\label{ee_inf}
\end{figure}

The presence of a systematic drift towards smaller values of $U_{c 1}$ as a
function of the system size may signal the presence of an intermediate phase
between the ordered and disordered ones. A first check on this hypothesis can
be obtained via investigation of the entanglement entropy. To this end, we
perform DMRG simulations up to $L=108$ sites. In Fig.~\eqref{ee_inf} we plot
the entanglement entropy for fixed $U = -1.95$ as a function of the cord
distance on the ring $\kappa(\ell) = L/ \pi \sin( \ell \pi / L )$, $\ell$ being
the length of the subsystem on the lattice, for different system sizes. By
directly fitting the scaling of the entropy for this value of $U$, which
belongs to the region between $U_{c1}$ and $U_{c2}$ according to all our
estimates, we are able to obtain a central charge in a good agreement with a
$c=1$ CFT. This is a strong indication of the presence of a critical phase for
$U>U_{c 1}$, compatible with the Luttinger liquid universality class.  We note
that for nonrelativistic critical points or phases, the entanglement entropy
is not bound by a logarithmic growth, and even if so, the coefficient could be
arbitrary. This implies that, assuming there is no fine-tuning, a $c = 1$ point
or phase is present here.

As discussed above, all entanglement-related quantities signal a single
second-order phase transition. This implies that the transition between the IC
and disordered phase shall belong to the BKT universality class, in agreement
with field theoretical insights~\cite{Chepiga2018,Fendley2004}.

By carrying out the same analysis of Sec.~\ref{procedure} on the lowest gap
in the energy spectrum, we can estimate the location of the BKT transition,
which is expected to occur for $U\ge-1.95$. The scaling ansatz differs from the
one in the previous section for two reasons: the dynamical critical exponent is
$z=1$, and the exponential divergence of the correlation length is
\be
\xi \sim \exp\( \frac{b}{\sqrt{U - U_{c 2 }}} \) ,
\ee
where $b$ is a nonuniversal constant, independent of $U$.  Moreover,
logarithmic corrections are known to intervene at the end of RG lines of fixed
points. In the case of a BKT point the functional form of these corrections is
known to be~\cite{BKTGAP1987} $\Delta \sim L^{-1} [ 1 + 1 /( 2 \ln L + C )
]^{-1}$, for some model-dependent constant $C$. On the basis of this field
theory result we take as the scaling function for the gap
\cite{Rigol2011,Dalmonte2015}
\be
\label{scal_bkt}
\Delta^* = L \(1+ \frac{1}{2 \ln L + C } \) \Delta = F \(  \frac{L}{\xi}  \) .
\ee
\noindent
This scaling ansatz, in combination with the procedure previously discussed,
has been tested in various spin chains where the location of the BKT transition
point was analytically known~\cite{Dalmonte2015}. In these cases, the method was
found to slightly underestimate the width of the gapless region; in our case
here, one thus expects that this method will overestimate the value of
$|U_{c2}|$. In terms of accuracy, the estimate obtained with this method is
compatible with state-of-the-art diagnostics based on targeting operator
dimensions via correlation functions.

\begin{figure}
{\includegraphics[scale=0.5]{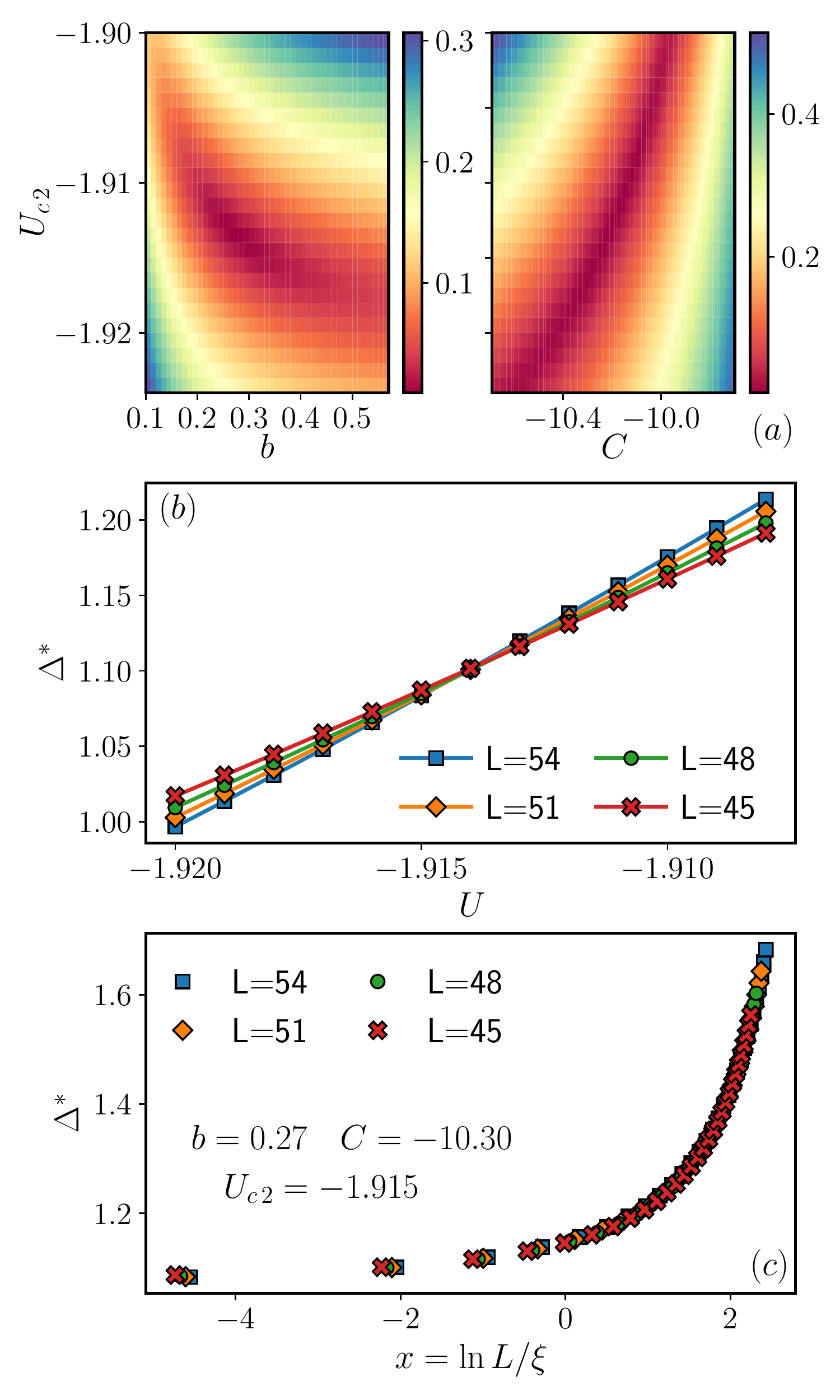}\hspace{2mm}}
\caption{(a) Density plot of the square root of the sum of the squared
  residuals in the $(b,U_{c1})$ and $(C,U_{c 1})$ planes for the best-fitting
  values of $C$ and $b$, respectively. (b) Crossing of the
  logarithmically corrected gaps, upon multiplication by $L^z$ (where $z = 1$)
  and taking the best-fit value for $C$. The crossing indicates the position of
  the critical point. (c) Data collapse of ED numerical data, with $U \in [U_{c
  2}, U_{c2} + 0.03]$ with the parameters $U_{c 2}, b,C$ which correspond to
  the best polynomial fit of the universal scaling function in
  Eq.~\eqref{scal_bkt}.}
\label{gap_bkt}
\end{figure}

In our case, we observe the same shifting of the critical point towards
negative $U$ as we take increasingly large system sizes. A sample result of the
largest system sizes we have investigated is plotted in Fig.~\ref{gap_bkt}; the
quality of the data collapse is excellent, as testified by the small value of
the sum of the discarded weights. By taking into account variations of the
optimal parameters with respect to the set of system sizes and amplitude of the
intervals considered, we get $b = 0.27 \pm 0.05$, $ C = 10.0\pm 0.5$, $U_{c 2}
= -1.915 \pm 0.008$ as the best estimate of the scaling function and transition
point location.

\section{Conclusions}
\label{sec5}

In this work, we have investigated the physics of the hard-core boson
constrained model of Eq.~\eqref{ham1} in the region of the phase diagram
surrounding the $\mathbb{Z}_3$-ordered phase. In the first part of the study,
we considered the vicinity of the Potts critical point. Since the position of
the latter is analytically known, we have used this regime to benchmark
entanglement-based techniques to detect quantum criticality in constrained
models. In particular, we have shown how concurrence and fidelity
susceptibility are able to accurately determine the exact location of the
critical point with accuracy of order $0.1\%$ in units of the coupling $U$. At
the critical point, we have carried out an extensive investigation of the
low-lying energy spectrum, matching such spectrum with the one expected from
the $\mathcal{M}_3$ minimal model. Our data suggest that it is possible, within
experimentally achievable system sizes, to unambiguously diagnose Potts quantum
criticality by just measuring spectral properties. We have also observed
systematic suppression of finite-size corrections in local observables, a
feature which we believe is due to integrability at the critical point.

In the second part of the work, we have investigated the melting of the ordered
phase in the so called doubly blockaded regime, that is, in the presence of
infinite next-to-nearest-neighbor repulsion. We have observed the presence of a
gapless regime, i.e., an incommensurate phase, already found in the same
model below the Potts transition point. Our results show how this phase is
surrounded by a second-order phase transition from one side, and a
Berezinskii-Kosterlitz-Thouless transition on the other. The position of the
latter has been determined using an advanced gap scaling technique at $U_{c 2}
= -1.915 \pm 0.008 $. 

Regarding the second-order phase transition, we have found that reaching a
scaling regime for entanglement (concurrence and fidelity susceptibility)
properties requires sizes $L>30$. Reaching this regime is also required to
determine the location of the transition point utilizing spectral properties.
Due to the difficulty in performing calculations for these sizes, entanglement
and spectral methods only allow us to provide a lower bound to the position of
this critical point, $U_{c1}\lesssim -1.96$. Similarly, we can only provide
bounds for the critical exponents; in particular, we find a systematic drift of
the value of $\nu$ toward smaller values, and of $z$ toward higher values.
These findings are not compatible with previous results~\cite{Samajdar2018}
based on sizes up to $L=36$, while they are compatible with a potential
emergence of Japaridze-Nerseyan-Pokrovsky-Talapov critical behavior observed below the Potts
point~\cite{Chepiga2018} and with a series of different non-relativistic
critical scenarios proposed in related field theories~\cite{Whitsitt:2018aa}.
Following the analogy between the FSS and the chiral clock model suggested in
Ref.~\onlinecite{Samajdar2018}, our findings indicate that, in the FSS model,
the critical line separating the ordered and disordered regimes ultimately
reaches the regime corresponding to large chiral angles in the clock model,
where an incommensurate phase intervenes between the two
phases~\cite{Zhuang:2015aa}.

We have complemented our analysis with numerical simulations monitoring the
behavior of solid order across the transition, using both quantum Monte Carlo
and tensor network methods. These methods predict a position of the phase
transition that strongly depends on the considered boundary conditions. In all
cases, the position of $U_{c2}$ is quite distinct from $U_{c1}$ with respect to the
numerical uncertainty of our results.

Our results suggest that the strong-coupling regime is relatively convenient to
observe phases with incommensurate order, as the size of the floating phase is
considerably larger than at smaller couplings. Moreover, spectral properties
should be favored as probes over correlation functions, which seem to be more
sensitive to finite-size effects. In addition, the presence of a relatively
extended transient scaling regime in terms of system sizes partly supports the
observation made in Ref.~\onlinecite{Ghosh:2018aa} regarding Kibble-Zurek
scaling: while the combined effects of a second-order and nearby BKT transition
have not been discussed in detail to the best of our knowledge, it is likely that
the presence of the latter affects rather dramatically the dynamics over
parameter space, due to exponentially vanishing gaps. We leave the
investigation of such a  scenario (which has been shown to be experimentally
achievable~\cite{Keesling:aa}) to a future study. Finally, it would be
interesting to systematically consider the effect of additional interaction
terms that are present in experiments: despite their modest magnitude (as they
decay very similarly to van der Waals interactions), those terms may sensibly
affect the size of the incommensurate phase.

\subsection*{Acknowledgements}

We acknowledge useful discussions with C. Castelnovo, R. Fazio, P. Fendley, V.
Khemani, C. Laumann, F. Mezzacapo, N. Prokof'ev, A. A. Nersesyan, N. Pancotti,
H. Pichler, L.  Piroli, S. Sachdev, and A. Sterdyniak.  This work is partly
supported by the ERC under Grant No. 758329 (AGEnTh), and has received funding
from the European Union Horizon 2020 research and innovation programme under
Grant agreement No. 817482." M.D. acknowledges computing resources at Cineca
Supercomputing Centre through the Italian SuperComputing Resource Allocation
via the ISCRA grant SkMagn. G.M. is partially supported through the project
``QUANTUM'' by Istituto Nazionale di Fisica Nucleare (INFN).

\appendix*
\section{MPS optimization in constrained Hilbert spaces}
\label{app}

We summarize our method to simulate constrained one-dimensional systems with tensor network techniques.
As opposed to local constraints or symmetries, which can be encoded directly at the level of the individual tensors, here we wish to take into account constraints spanning several neighboring sites.
In the context of this type of spin model, a recent approach has been
implemented in Ref.~\onlinecite{Chepiga2018}, by keeping track of the possible
transitions between valid states when constructing the environment tensors, for
the case of the NN blockade.
Our method instead aims at being slightly more general, in order to accommodate arbitrary nonlocal constraints --- and, in particular, NNN ones --- with a low computational overhead.
More specifically, we encode a state in the full many-body Hilbert space, and then construct the projector to the specific subspace we are interested in. 

For this paper, we consider the Hilbert space $\H_L = (\mathbb{C}^2)^{\otimes L}$, where we label the local basis by $\{\ket{\circ},\ket{\bullet}\}$.
States in $\H_L$ can be represented by MPS with a local physical dimension $d=2$.  
\begin{figure}
  \centerline{
    \begin{tikzpicture}[auto,every node/.style={scale=0.8}]
    \node[state] (1) {\texttt{1}};
    \node[state,right=of 1] (2) {\texttt{2}};
    \path[->] (1) edge[bend left] node[above]  {$n$} (2);
    \path[->] (2) edge[bend left] node[below]  {$\tilde{n}$} (1);
    \path[->] (1) edge[loop below] node[below]  {$\tilde{n}$} (1);
  \end{tikzpicture}
  \hspace{20pt}
  \begin{tikzpicture}[auto,every node/.style={scale=0.8}]
    \node[state] (1) {\texttt{1}};
    \node[state,right=of 1] (2) {\texttt{2}};
    \node[state,right=of 2] (3) {\texttt{3}};
    \path[->] (1) edge[bend left] node[above]  {$n$} (2);
    \path[->] (2) edge[bend left] node[above]  {$\tilde{n}$} (3);
    \path[->] (3) edge[bend left] node[below]  {$\tilde{n}$} (1);
    \path[->] (1) edge[loop below] node[below]  {$\tilde{n}$} (1);
  \end{tikzpicture}
  }
  \caption{\label{fig:auto} Finite-state automaton corresponding to the nearest-neighbor ({\it left}) and next-nearest-neighbor ({\it right}) projector. Notice that there is not a clear initial and final state, as in the case of Hamiltonians, but these automata form a cycle.}
\end{figure}
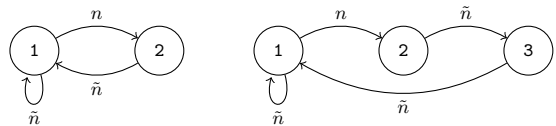
We can then construct the projector $\mathcal{P}$ onto the subspace of interest as an MPO.
The construction of an arbitrary operator can be achieved by exploiting the
correspondence between MPOs and finite-state automata~\cite{Crosswhite2008}. We
label the internal states as \texttt{1},\texttt{2},\dots. In the case of the
single constraint $n_i n_{i+1} = 0$, the only nonzero entries on each tensor
of the projector are 
\be
  {
  \tikz[baseline=-0.5*\tu,node distance=\tu]{
      \node[mps]  (m) {};
      \draw[index] (m.north) -- +(0,\tu) node[above]{$\circ$};
      \draw[index] (m.south) -- +(0,-\tu) node[below]{$\circ$};
      \draw[index] (m.west) -- +(-\tu,0) node[left]{\texttt{1}};
      \draw[index] (m.east) -- +(\tu,0) node[right]{\texttt{1}};
    }
  } = 
  {
  \tikz[baseline=-0.5*\tu,node distance=\tu]{
      \node[mps]  (m) {};
      \draw[index] (m.north) -- +(0,\tu) node[above]{$\bullet$};
      \draw[index] (m.south) -- +(0,-\tu) node[below]{$\bullet$};
      \draw[index] (m.west) -- +(-\tu,0) node[left]{\texttt{1}};
      \draw[index] (m.east) -- +(\tu,0) node[right]{\texttt{2}};
    }
  } =
  {
  \tikz[baseline=-0.5*\tu,node distance=\tu]{
      \node[mps]  (m) {};
      \draw[index] (m.north) -- +(0,\tu) node[above]{$\circ$};
      \draw[index] (m.south) -- +(0,-\tu) node[below]{$\circ$};
      \draw[index] (m.west) -- +(-\tu,0) node[left]{\texttt{2}};
      \draw[index] (m.east) -- +(\tu,0) node[right]{\texttt{1}};
    }
  }
  = 1
\ee
which corresponds to the finite-state automaton on the left of Fig.~\ref{fig:auto}.
The corresponding MPO tensor for the projector is
\be
  P_{a b} =
  \begin{pmatrix}
  \tilde{n} & n \\
  \tilde{n} & 0 
  \end{pmatrix}  
\ee
where $\tilde{n} = \ket{\circ}\bra{\circ}$ and $n = \ket{\bullet}\bra{\bullet}$.
Similarly, we can construct the projector with a next-nearest-neighbor interaction; i.e., we want to simultaneously impose $n_i n_{i+1} = 0$ and $n_i n_{i+2} = 0$.
We modify the previous approach and add the diagrams
\be
  {
  \tikz[baseline=-0.5*\tu,node distance=\tu]{
      \node[mps]  (m) {};
      \draw[index] (m.north) -- +(0,\tu) node[above]{$\circ$};
      \draw[index] (m.south) -- +(0,-\tu) node[below]{$\circ$};
      \draw[index] (m.west) -- +(-\tu,0) node[left]{\texttt{1}};
      \draw[index] (m.east) -- +(\tu,0) node[right]{\texttt{1}};
    }
  } =
  {
  \tikz[baseline=-0.5*\tu,node distance=\tu]{
      \node[mps]  (m) {};
      \draw[index] (m.north) -- +(0,\tu) node[above]{$\bullet$};
      \draw[index] (m.south) -- +(0,-\tu) node[below]{$\bullet$};
      \draw[index] (m.west) -- +(-\tu,0) node[left]{\texttt{1}};
      \draw[index] (m.east) -- +(\tu,0) node[right]{\texttt{2}};
    }
  } =
  {
  \tikz[baseline=-0.5*\tu,node distance=\tu]{
      \node[mps]  (m) {};
      \draw[index] (m.north) -- +(0,\tu) node[above]{$\circ$};
      \draw[index] (m.south) -- +(0,-\tu) node[below]{$\circ$};
      \draw[index] (m.west) -- +(-\tu,0) node[left]{\texttt{2}};
      \draw[index] (m.east) -- +(\tu,0) node[right]{\texttt{3}};
    }
  } =
  {
  \tikz[baseline=-0.5*\tu,node distance=\tu]{
      \node[mps]  (m) {};
      \draw[index] (m.north) -- +(0,\tu) node[above]{$\circ$};
      \draw[index] (m.south) -- +(0,-\tu) node[below]{$\circ$};
      \draw[index] (m.west) -- +(-\tu,0) node[left]{\texttt{3}};
      \draw[index] (m.east) -- +(\tu,0) node[right]{\texttt{1}};
    }
  }
  = 1
\ee
This corresponds to the automaton on the right of Fig.~\ref{fig:auto} and a MPO tensor
\be
  P_{a b} =
  \begin{pmatrix}
  \tilde{n} & n & 0  \\
  0 & 0 & \tilde{n} \\
  \tilde{n} & 0 & 0 
  \end{pmatrix}  
\ee

To approximate the ground state $\ket{\psi}$ of a certain Hamiltonian $H$ under the constraint, we can then proceed by variationally updating single tensors such that they minimize the projected Hamiltonian $\langle \tilde{H} \rangle = \brakett{\psi}{\mathcal{P} H \mathcal{P}}{\psi}$.
The optimization algorithm will automatically converge towards a state that satisfies the constraint.
This can be understood by decomposing the state into parallel and perpendicular components to the subspace satisfying the constraint: $\ket{\psi} = \alpha \ket{\psi_\parallel} + \beta \ket{\psi_\bot}$, with the usual normalization $\alpha^2 + \beta^2 = 1$.
The energy expectation value is then $\brakett{\psi}{\mathcal{P} H \mathcal{P}}{\psi} = \alpha^2 \brakett{\psi_\parallel}{H}{ \psi_\parallel}$.
Any state in the perpendicular subspace will be an eigenvector of $\tilde{H}$, with eigenvalue $0$. 
Hence, if the true ground-state energy is negative, the optimization algorithm naturally favors the parallel part, sending $\alpha \to 1$, $\beta \to 0$.
This condition can always be fulfilled without loss of generality, since we can always shift the Hamiltonian by a constant to ensure $\brakett{\psi_\parallel}{H}{ \psi_\parallel} < 0$.
For the situations investigated here, we observe that this procedure works
quite well, and the constraint violation $(1 - |\langle \mathcal{P}
\rangle|^2)/L$ is consistent with the order of magnitude of machine precision. 

In order to speed up the optimization, it is natural to contract $\tilde{H}$
into a single MPO. For the purely local Hamiltonian, as in Eq.~\eqref{ham1},
the bond dimension is 2. The bond dimension of these MPOs can be reduced down
to $4$ (single constraint) and $6$ (double constraint) by removing identical
rows\cite{Claudius2017}. This can be also checked by performing a singular
value decomposition of the MPO\cite{Schol2011}. As an additional improvement,
it is convenient to block two or three sites together into a single tensor,
depending on the projector. By imposing the constraint on this new block, the
number of local physical states is $d=3$ in the case of the single constraint
with a 2-sites block and $d=4$ in the case of the double constraint with a
3-sites block.

\phantomsection
\addcontentsline{toc}{chapter}{Bibliography}

\bibliography{bib-GG}{}

\end{document}